\documentclass[notitlepage,aps,reprint,superscriptaddress,nofootinbib,prd]{revtex4-1}
\usepackage{epsfig,verbatim}
\usepackage{bbold}
\usepackage{dsfont}
\usepackage[caption=false]{subfig}
\usepackage{amsmath,amssymb,graphicx}
    \graphicspath{ {./figures/} }
\usepackage{color}
\usepackage{slashed}            
\usepackage{bbm}                
\usepackage{units}              
\usepackage{hyperref}           
\usepackage{xspace}             
\usepackage{enumerate}          
\usepackage{soul}
\usepackage{cancel}
\usepackage{dcolumn}
\usepackage{mathalfa}

\usepackage{amstext}    
\usepackage{array} 
\usepackage{dcolumn}   
\usepackage{bm}        
\usepackage{amssymb,amsthm}   
\usepackage{mathrsfs}
\usepackage[british]{babel}
\usepackage[applemac]{inputenc}
 \usepackage{multirow}
\usepackage{setspace}

\usepackage{empheq}

\usepackage{bbm}                
\usepackage{xspace}             
\usepackage{upgreek}
    
\newcolumntype{L}[1]{>{\raggedright\let\newline\\\arraybackslash\hspace{0pt}}m{#1}}
\newcolumntype{C}[1]{>{\centering\let\newline\\\arraybackslash\hspace{0pt}}m{#1}}
\newcolumntype{R}[1]{>{\raggedleft\let\newline\\\arraybackslash\hspace{0pt}}m{#1}}

\newcommand{\rme}{\mathrm{e}}
\newcommand{\rmi}{\mathrm{i}}
\newcommand{\be}{\begin{equation}}
\newcommand{\ee}{\end{equation}}
\newcommand{\bea}{\begin{eqnarray}} 
\newcommand{\eea}{\end{eqnarray}}
\newcommand{\ft}[2]{{\textstyle\frac{#1}{#2}}}

\renewcommand\[{\left[}
\renewcommand\]{\right]}
\renewcommand{\Re}{\operatorname{Re}}
\renewcommand{\Im}{\operatorname{Im}}

\newcommand{\Sp}{\mathop{\rm {}Sp}}

\def\cF{{\cal F}}
\def\cG{{\cal G}}
\def\cN{{\cal N}}
\def\cS{{\cal S}}
\def\cV{{\cal V}}
\def\pd{\ensuremath{\partial}}

\usepackage[dvipsnames]{xcolor}

\begin{document}

\title{Universal Class of Type-IIB Flux Vacua with Analytic Mass Spectrum}

\author{Jose J. Blanco-Pillado}
\email[]{josejuan.blanco@ehu.eus}
\affiliation{Department of  Physics, University of the Basque Country UPV/EHU, 48080 Bilbao, Spain}
\affiliation{IKERBASQUE, Basque Foundation for Science, 48011, Bilbao, Spain}
\author{Kepa Sousa}
\email[]{kepa.sousa@utf.mff.cuni.cz}
\affiliation{Institute of Theoretical Physics, Charles University, V Holes\v{o}vi\v{c}k\'ach 2, Prague, Czech Republic}
\author{Mikel A. Urkiola}
\email[]{mikel.alvarezu@ehu.eus}
\affiliation{Department of  Physics, University of the Basque Country UPV/EHU, 48080 Bilbao, Spain}
\author{Jeremy M. Wachter.}
\email[]{jwachter@skidmore.edu}
\affiliation{Skidmore College Physics Department, 815 North Broadway Saratoga Springs, New York 12866}

\date{\today}

\begin{abstract}
We report on a new class of flux vacua generically present in Calabi-Yau compactifications 
of type-IIB string theory. At  these vacua, the mass spectrum of the complete axio-dilaton/complex 
structure sector is given,  to leading order in $\alpha'$ and $g_s$, by a simple analytic formula 
independent of the choice of Calabi-Yau. We provide a method to find these vacua, and 
construct an ensemble of 17,054 solutions for the Calabi-Yau hypersurface 
$\mathbb{WP}_{[1,1,1,6,9]}^4$, where the masses of the axio-dilaton and the 
$272$ complex structure fields can be explicitly computed.
\end{abstract}

\maketitle


\section{Introduction}

The study of the phenomenological implications of string theory demands the construction of low-energy Effective Field Theories (EFTs) describing  its compactification to four dimensions. However, deriving  these EFTs is remarkably challenging and involves, in particular, integrating out a large number of scalar fields (typically hundreds) describing the geometry of the compactified dimensions, i.e., the \emph{moduli}. Actually, finding a mechanism to generate the moduli masses is a crucial step in the best studied proposals to construct de Sitter vacua in type-IIB string theory, i.e., the KKLT  \cite{Kachru:2003aw}  and Large Volume Scenarios (LVS)  \cite{Balasubramanian:2005zx,Conlon:2005ki}. Both constructions rely on the identification of \emph{flux vacua}: minima of the effective potential induced by higher-dimensional form fields. Although the general features of the flux potential are  well characterised, a detailed computation of the moduli mass spectra in generic scenarios is still extremely difficult due to the complexity of the theory and the large number of fields involved. More specifically, the so-called \emph{no-scale structure} of the flux potential  ensures that a subset of the moduli, namely the axio-dilaton and complex structure fields, can be fixed at a perturbatively stable configuration provided they preserve supersymmetry. 

However, the stabilisation of the remaining moduli fields, i.e., the  K\"ahler moduli, requires including $\alpha'$ perturbative corrections   and nonperturbative contributions which spoil the no-scale structure \cite{Gross:1986iv,Becker:2002nn} (see also \cite{Sethi:2017phn,Anguelova:2010ed,Kachru:2018aqn}). Therefore   the uncorrected  flux vacua may become tachyonic, or even cease being critical points of the potential. Indeed, while the stability of the axio-dilaton/complex structure sector in  the fully stabilised vacuum has been argued in KKLT and LVS  using scaling arguments \cite{Abe:2006xp,Gallego:2008qi,Gallego:2009px,Abe:2006xp,Conlon:2005ki,Rummel:2011cd}  and by the direct examination of explicit examples (see, e.g., \cite{Louis:2012nb,Demirtas:2020ffz}), it has rarely been studied in detail.  Actually, as discussed in \cite{Conlon:2008,Achucarro:2015kja,Sousa:2014qza},  in both KKLT and LVS scenarios the presence of light (or massless) fields in the spectrum to leading order in $\alpha'$ and quantum corrections may still lead to the appearance of instabilities in the final vacuum. Interestingly, the presence of such dangerously light modes has been reported to arise in explicit constructions of de Sitter vacua, where the moduli are stabilised near special points of the moduli space \cite{Hebecker:2006bn,Bena:2018fqc,Demirtas:2019sip}. Furthermore, recent analyses indicate an existing tension between the $D3-$tadpole cancellation condition and the need to stabilise all the complex structure moduli \cite{Braun:2020jrx,Crino:2020qwk,Bena:2020xrh,Bena:2021wyr}, what could have important implications for the consistency of KKLT and LVS proposals. While significant progress has been made in crucial aspects of moduli stabilisation in the last couple of years\cite{Demirtas:2019sip,Demirtas:2020ffz,Blumenhagen:2020ire,Crino:2020qwk}, a precise characterisation of the mass spectrum in the axio-dilaton/complex structure sector has remained elusive, which calls for further studies in this direction.

Advances on this matter were recently  made in  \cite{Blanco-Pillado:2020wjn} (following \cite{Sousa:2014qza,Brodie:2015kza,Marsh:2015zoa}), for compactifications on the orientifold of  Calabi-Yau manifolds which allow the consistent truncation of all the complex structure fields except one. The analysis of \cite{Blanco-Pillado:2020wjn}  assumed a Calabi-Yau geometry admitting a large discrete isometry group which, provided the flux configuration is also invariant under these symmetries, allows the effective reduction of the complex structure sector. In this setting it was shown that \emph{the complete mass spectrum} of  the axio-dilaton and complex structure sector (including the truncated fields) can be explicitly computed in the  Large Complex Structure (LCS)/weak string coupling regime. More specifically, for the class of vacua which can be found parametrically close to the LCS point \cite{Junghans:2018gdb,Grimm:2019ixq} and up to exponentially small corrections, the scalar moduli masses are given by the simple analytic formula
\be
\frac{\mu_{\pm\lambda}^2}{m_{3/2}^2} =
\left\{
 \begin{array}{l c l }
\left( 1 \pm \frac{\sqrt{(1-2 \xi)}}{\sqrt{3}} \hat m(\xi) \right)^2&~&  \hspace{-.1cm}\lambda=0,\\
\left( 1 \pm \frac{\sqrt{(1-2 \xi)}}{\sqrt{3}\hat m(\xi)} \right)^2&~& \hspace{-.1cm} \lambda=1,\\
\left(1 \pm \frac{1+\xi}{3}\right)^2 &~& \hspace{-.1cm}\lambda=2,\ldots, h^{2,1}_-.
\end{array}
\right.\,
\label{eq:N0Spectrum}
\ee
Here, the quantity $\xi$ parametrises  the complex structure moduli space, ranging in $\xi \in [0,1/2)$ for $h^{2,1}_->h^{1,1}_+$ or in $\xi \in (-1,0]$ for $h^{2,1}_-<h^{1,1}_+$, with the LCS point located at $\xi=0$, and with  $h^{2,1}_-$ and $h^{1,1}_+$ denoting the number of complex structure and K\"ahler moduli fields of the Calabi-Yau orientifold, respectively. The quantity $m_{3/2}$ is the gravitino mass, and 
\be
\hat m(\xi) \equiv \ft{1}{\sqrt{2}}\left(2 + \kappa(\xi)^2 - \kappa(\xi) \sqrt{4 + \kappa(\xi)^2}\right)^{1/2}
\label{eq:mhat}
\ee
with $\kappa(\xi) = 2(1+\xi)^2 / \sqrt{3(1 - 2\xi)^3}$.

In the present paper we prove that, in the LCS regime and provided the fluxes are conveniently constrained, the EFT of \emph{generic Calabi-Yau compactifications} always admits a consistent truncation of all complex structure fields but one.    Here, in contrast with \cite{Blanco-Pillado:2020wjn},  we do not  require the presence of a discrete isometry group, and our result relies instead on the existence of monodromy transformations around the LCS point. That is, we will require only the invariance of the EFT under discrete shifts of the complex structure fields $z^i$
\be
z^i \to z^i +  v^i,  \qquad v^i \in \mathbb{Z}^{h^{2,1}}, 
\label{eq:monodromy}
\ee
combined with an appropriate  transformation of the fluxes of the form fields. This invariance is a common feature of all Calabi-Yau compactifications in the LCS regime. 
Moreover, the choice of  field surviving the truncation is highly non-unique, with each possibility associated to a different monodromy direction $v^i$.

This simple, and yet powerful, observation allows us to extend the  results of \cite{Blanco-Pillado:2020wjn} to generic Calabi-Yau compactifications and, as a consequence, opens the door to generating a large landscape of vacua with  an unprecedented analytic control over the mass spectrum of the axio-dilaton and  complex structure moduli. It is important to note that solutions discussed here  are not the dominant class in the landscape. In fact, they only constitute a small fraction of the total number of vacua for large values of $h^{2,1}_-\gg1$. However, as we shall see below, our approach allows us to search for these solutions in a very efficient way, which makes it particularly attractive for explicit constructions of de Sitter vacua.

\section{Consistent truncation of the EFT}

We begin by introducing the effective supergravity  theory describing  the low-energy regime of type-IIB string theory compactified on a Calabi-Yau orientifold $\tilde M_3$. 
The couplings of the theory are conveniently expressed by specifying an integral and symplectic homology basis $\{A^I,B_I\}$ of $H_3(X_3,\mathbb{Z})$, satisfying $A^I\cap B_J =\delta^I_J$ and $A^I \cap A^J = B_I \cap B_J =0$,
with $I=0, \ldots, h^{2,1}_-$. In particular, the components in this basis of the Calabi-Yau $(3,0)$ form  $\Omega(z^i)$  can be encoded in the  \emph{period vector}
\be
\Pi^T \equiv(\cF_I,X^I)= \begin{pmatrix}
\int_{B_I} \Omega, &  \int_{A^I} \Omega 
\end{pmatrix}.
\ee
Here $X^I$ are projective coordinates in the complex structure moduli space, and the corresponding moduli fields can 
be defined to be $z^i \equiv X^i/X^0$, $i = 1, \ldots, h^{2,1}_-$.   Then, to leading order in $\alpha'$ and the string coupling $g_s$, the K\"ahler potential  of the corresponding 4-dimensional effective supergravity theory reads  
\cite{Grimm:2004uq,Klemm:2005tw}
\be
K = - 2 \log \cV 
- \log(-\rmi (\tau-\bar \tau)) - \log \left(  -\rmi \, \Pi^\dag \cdot \Sigma\cdot \Pi\,\right)\,, 
\label{eq:KahlerPotential}
\ee
where $\cV$ denotes the K\"ahler moduli-dependent volume of $\tilde M_3$  in units of $2\pi \sqrt{\alpha'}$ and 
measured  in Einstein frame, and   $\Sigma = \tiny{\begin{pmatrix}
0 & \mathbb{1}\\
-\mathbb{1} & 0
\end{pmatrix}}$ is the symplectic matrix.
The quantities $\cF_I$ can be expressed as the derivatives of a holomorphic 
function of the $X^I$, the \emph{prepotential} $\cF(X^I)$, which
in the LCS regime admits the expansion 
\be
\cF = -\ft{1}{3!} \kappa_{ijk} z^i z^j z^k - \ft{1}{2!} \kappa_{ij} z^i z^j +\kappa_i z^i + \ft{1}{2} \kappa_0 \, +\ldots\; ,
\label{eq:F}
\ee
where we have chosen the gauge $X^0 =1$.
The terms $\kappa_{ijk}$, $\kappa_{ij}$ and $\kappa_i$ are numerical constants which can be computed from the 
topological data of the mirror manifold to $M_3$ (see \cite{Hosono:1994ax}). In particular, the quantities $\kappa_{ijk}$  are integers, 
the coefficients $\kappa_{ij}$ and $\kappa_i$ are rational, and the constant $\kappa_0=\zeta(3)\chi(M_3)/(2 \pi \rmi)^3$ is determined by the Euler number $\chi(M_3)$ of the Calabi-Yau.  
 The prepotential also receives contributions from  world-sheet instantons,  which are subleading in the LCS regime, and thus they will  be neglected
 in the following calculations.
The presence of the RR and NS-NS three-form fluxes, resp. $F_{(3)}$ and   $H_{(3)}$, 
 induces the Gukov-Vafa-Witten superpotential $W$ for the dilaton and complex structure moduli \cite{Gukov:1999ya}
\be
 \sqrt{\pi/2} \,  W=  N^T \cdot \Sigma\cdot \Pi ,
\label{eq:EFT}
\ee
where we have  introduced the flux vector 
\be
N\equiv f-\tau h,\quad  f= 
\begin{pmatrix}
\int_{B^I} F_{(3)}\\
\int_{A_I} F_{(3)}
\end{pmatrix}, \quad h=  
\begin{pmatrix}
\int_{B^I} H_{(3)}\\
\int_{A_I} H_{(3)}
\end{pmatrix},
\label{eq:fluxes}
\ee
with $\{f_{A}^I, f^B_I, h_{A}^I, h^B_I\}\in \mathbb{Z}$. Then, the configurations of the axio-dilaton and complex structure fields $\{\tau_c, z^i_c\}$ which  minimize the scalar potential while preserving supersymmetry  are those 
satisfying the $F$-flatness conditions
\be
(\pd_\tau  + \pd_\tau K) W|_{\tau_c, z^i_c} = (\pd_{z^i}  + \pd_{z^i} K) W|_{\tau_c, z^i_c} =0.
\label{eq:noScale}
\ee
The present description of the EFT  has an inherent redundancy associated to the choice of homology basis. More specifically,  a change of basis  induces a transformation of the period and flux vectors, 
\be
\Pi \to \cS \cdot \Pi, \qquad N \to \cS \cdot N, 
\label{eq:symplecticTrans}
\ee
with $\cS \in \Sp(2h^{2,1}_-+2,\mathbb{Z})$, leading to different descriptions of the same theory. 
Finally, the requirement that the period vector transforms by symplectic transformations under the monodromies \eqref{eq:monodromy} leads to the following  condition on the couplings \cite{Candelas:1993dm,Mayr:2000as}
\be
\kappa_{ij} v^j + \ft12  \kappa_{ijk} v^j v^k =0 \mod \mathbb{Z}.
\label{eq:couplingConstraint}
\ee
We will now prove the main result of this paper:

\emph{Let us consider  a  $h^{2,1}_-$-dimensional vector $v^i$ of coprime integers, which lies in the K\"ahler cone of the mirror Calabi-Yau.  
Then,  the ansatz $z^i = \hat z v^i$ with $\hat z \in \mathbb{C}$ defines a consistent supersymmetric truncation of the EFT given by  \eqref{eq:KahlerPotential}, \eqref{eq:F}   and  \eqref{eq:EFT} when the 
flux  configuration is of the form 
\bea
N_A^0&=&0,\qquad 
 N_A^i= v^i \hat N_A, \nonumber \\
 N^B_i &=& q\kappa_{ijk} v^j v^k \hat N^B - (\kappa_{i j} + \ft12\kappa_{ijk}v^k ) N_A^j,
\label{eq:constrainedFluxes}
\eea
and $N_0^B$ arbitrary. Here  $\hat N_A \equiv \hat f_A - \tau \hat h_A$, $\hat N^B \equiv \hat f ^B - \tau \hat h^B$ with $\{\hat f_A, \hat h_A, \hat f ^B,\hat h^B\} \in\mathbb{Z}$ and $q^{-1} \equiv gcd(\kappa_{ijk} v^j v^k)$. }

\emph{Proof.} First, note that the constraint  \eqref{eq:couplingConstraint} ensures that   the vectors $f$ and $h$ defined in \eqref{eq:fluxes} have  integer components, as required by the flux  quantization condition.
 To prove that the ansatz  $z^i =\hat  z v^i$ with $\hat z \in \mathbb{C}$ defines a consistent supersymmetric truncation of the EFT with the fluxes \eqref{eq:constrainedFluxes}, we need to check that the $F$-flatness condition, $w^i (\pd_{z^i}+\pd_{z^i}K)W|_{\hat z v^i}=0$, is satisfied along all directions $w^i$ orthogonal to the reduced field space defined by the truncation ansatz, i.e., orthogonal to $v^i$, regardless  of the value of $\hat z$ and $\tau$ \cite{Achucarro:2007qa,Achucarro:2008sy,Achucarro:2008fk,Sousa:2014qza}.
Substituting the  flux configuration \eqref{eq:constrainedFluxes} into \eqref{eq:EFT} we find that the $F$-flatness condition reads
\be
 \kappa_{ijk}  w^i v^j v^k [(\hat z +\ft{\rmi}{2}) N_A- \rmi q N^B]+ w^i [\pd_{z^i} K\, W]_{\hat z v^i} =0.
\label{eq:truncation}
\ee
Actually the two terms in this expression vanish independently, as they are both proportional to $\kappa_{ijk}\, w^i \Im(z^j) \Im( z^k)  = 0$,
which is zero in the  LCS regime. Indeed, this quantity vanishes at any configuration  $z^i$ where the holomorphic vector $w^i$ is orthogonal to $\Im(z^i)$ \cite{Blanco-Pillado:2020wjn} (see also  appendix \ref{app:Fcondition} and \cite{Cremmer:1984hj,Farquet:2012cs,Marsh:2015zoa}). \hspace{.5cm}  $\blacksquare$ \\

The previous result guarantees that  the  ansatz $z^i = \hat z v^i$  can be consistently substituted into the   action,  obtaining a reduced  theory with an effectively $1$-dimensional complex structure moduli space parametrised by $\hat z$. The couplings of the reduced action are still characterised  by \eqref{eq:KahlerPotential} and \eqref{eq:EFT}, but with  an effective prepotential given by
\be
\hat \cF\equiv -\ft{1}{3!} \kappa_{vvv} \hat z^3  + \ft{1}{2 \cdot 2!}  \kappa_{vvv} \hat z^2 +\kappa_v \hat z + \ft{1}{2} \kappa_0
\label{eq:effPrep}
\ee
and an effective $4$-dimensional flux vector 
\be
\hat  N  \equiv ( N^B_0, \, q\kappa_{vvv} \hat N_B, \,  0,\,  \hat N_A)^T,
\label{eq:effFlux}
\ee
where we introduced the shorthands\footnote{The freedom \eqref{eq:symplecticTrans} allows to shift $\kappa_{ij} v^iv^j$ by an arbitrary integer, what we use to set $\kappa_{ij} v^iv^j = -\ft12 \kappa_{ijk} v^i v^j v^k$ in \eqref{eq:effPrep} \cite{Candelas:1993dm}.} $\kappa_{vvv} \equiv \kappa_{ijk} v^i v^j v^k$ and $\kappa_v \equiv \kappa_i v^i$.
Any solution of this reduced theory is also a solution of the full action in the LCS regime and to leading order in $\alpha'$ and $g_s$. Furthermore,  if the fields surviving the truncation  satisfy the $F$-flatness conditions \eqref{eq:noScale},  the  axio-dilaton/complex structure sector of the complete theory will satisfy them  as well  \cite{Sousa:2012nvn,Sousa:2014qza}. 

Therefore, given an EFT for some Calabi-Yau compactification, we can immediately generate large families of flux vacua in the LCS regime (one family for each choice of $v^i$)  where we can compute the mass spectrum of the complete axio-dilaton/complex structure sector. Indeed,  we just need solve the $F$-flatness  conditions \eqref{eq:noScale} for the reduced model defined by \eqref{eq:effPrep} and \eqref{eq:effFlux}. Then, the mass spectrum at the resulting vacua can be obtained using the results in \cite{Blanco-Pillado:2020wjn}, which  apply whenever the complex structure sector can be consistently truncated to a single field.
More specifically,  the formula  \eqref{eq:N0Spectrum} gives the squared masses of all the $2 h^{2,1}_- +2$ scalar modes in the  axio-dilaton/complex structure sector, including the truncated ones,  in terms of a single  parameter $\xi \equiv  \frac{-3 \Im \kappa_0}{2 \kappa_{vvv} \Im(\hat z)^3}$, and  normalised by the gravitino mass, 
\be
m_{3/2}^2 \equiv \rme^{K} |W|^2=3  Q_{D3} /(\pi(2-\xi)\cV^2), 
\label{eq:gravitino}
\ee
where $Q_{D3}\equiv f^T \cdot  \Sigma \cdot h\ge1$ is the flux induced $D3-$charge.
 In equation \eqref{eq:N0Spectrum}, the masses with $\lambda =0,1$ are those associated to the fields surviving the truncation, $\{\tau, \hat z\}$, while those with $\lambda = 2, \ldots, h^{2,1}_-$ are the masses of the remaining fields  in the truncated sector.  It is also worth mentioning that   solutions to \eqref{eq:noScale}  with $N_A^0=0$ are of particular interest, as they are the only ones that can be found  parametrically close to the  LCS point \cite{Brodie:2015kza,Marsh:2015zoa,Junghans:2018gdb,Grimm:2019ixq,Blanco-Pillado:2020wjn},  which is where we have the best perturbative control of the EFT. 

To end this section, let us briefly comment on the $D3$-tadpole constraint. In a given compactification, the number of solutions at LCS compatible with the spectrum \eqref{eq:N0Spectrum}, $\cN$, can be estimated using the continuous flux approximation of \cite{Denef:2004ze}. We find
\be
\cN(Q_{D3}\le Q_{D3}^*,g_s\le g_s^*) \propto \sum_{v^i \in \mathrm{CK}} \frac{g_s^*\, |\Im \kappa_0| (Q_{D3}^*)^3}{q^2 \kappa_{vvv}^2},
\label{eq:Nvac}
\ee
where $Q_{D3}^*$ is the available $D3$-charge in the compactification, $g_s = (\Im \tau)^{-1}$ is the string coupling, and the proportionality constant is of order one (see appendix \ref{app:Nvac}). The sum in the previous formula extends over all vectors $v^i$ in the K\"ahler cone (CK) of the mirror dual to $M_3$. Although the actual number of vacua depends on the choice of compactification,   this result shows  that $\cN$ only represents a very small fraction of the total number of flux vacua  $\cN \ll \cN_{\text{total}} \propto  (Q_{D3}^*)^{2 (h^{2,1}_-+1)}$  when  $h^{2,1}_-\gg1$ \cite{Denef:2004ze}. Nevertheless, the method described above allows us to very efficiently search for these solutions, as we demonstrate next with an explicit example.

\section{Example: the hypersurface $\mathbb{WP}_{[1,1,1,6,9]}^4$}

 We  will now illustrate our results by constructing an ensemble of the class  of vacua presented above. For this purpose we will consider    the compactification of type-IIB string theory in an orientifold of the  Calabi-Yau hypersurface $\mathbb{WP}_{[1,1,1,6,9]}^4$, which has $h^{1,1}_+=2$ K\"ahler moduli and  and $h^{2,1}_- = 272$ complex structure fields.  For geometries admitting  a $\cG=\mathbb{Z}_{18}\times \mathbb{Z}_6$  isometry group, and provided only $\cG$-invariant fluxes are turned on, the complex structure sector can be consistently truncated, leaving only two surviving complex structure fields which also transform trivially under $\cG$. In the LCS regime, the couplings for the two $\cG$-invariant complex structure fields  are determined by a prepotential with coefficients \cite{Candelas:1994hw}
\bea
\kappa_{111} = 9, \quad \kappa_{112} &=& 3, \quad \kappa_{122} =1,\nonumber \\
\kappa_{11} = -\ft{9}2, \quad\kappa_{22}&=&0,\quad  \kappa_{12} = -\ft32, 
\label{eq:couplings}
\eea
$\kappa_i =( \ft{17}{4},\,  \ft32)$ and $\kappa_0= -540 \zeta(3) /(2 \pi \rmi )^3$. Recall that the presence of the group $\cG$ is not necessary for our results to apply, however such isometries are often required  to make the computation of the  EFT couplings tractable  (see  \cite{Cicoli:2013cha}). We will also assume the same configuration of orientifold planes and $D7$-branes as in \cite{Demirtas:2019sip}, which allows for flux vacua with  $D3$-charge  satisfying $Q_{D3} \le 138$.

\begin{figure}[t]
    \centering 
	\hspace{-.4cm}\includegraphics[width=0.47\textwidth]{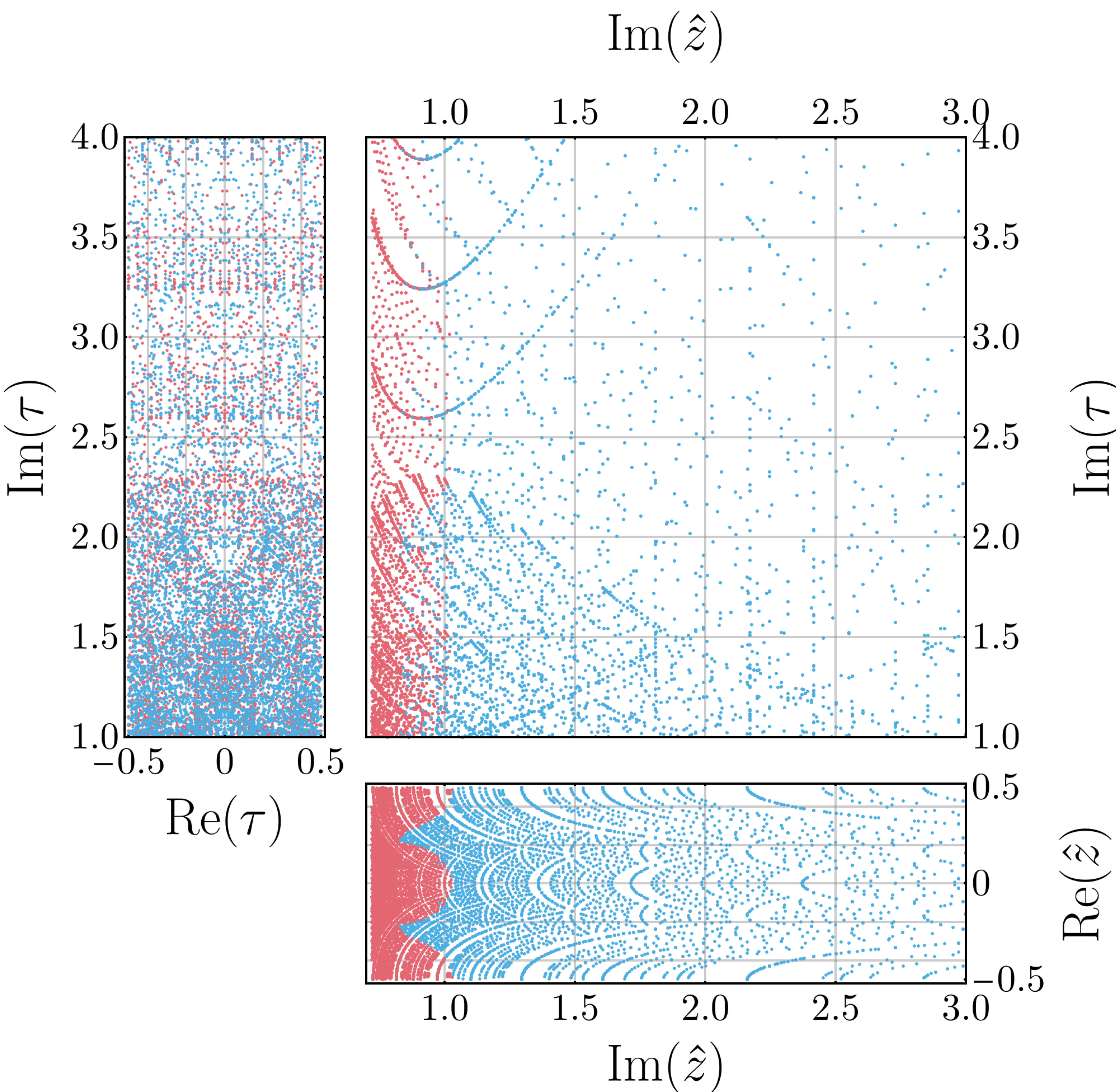} 
    \caption{Numerically generated distribution of flux vacua  on the fundamental domain of the reduced  field space, with $\Re \tau \in [-1/2, 1/2)$, $\Im \tau>1$ and $\Re \hat z \in [-1/2,1/2)$,  for the $\mathbb{WP}_{[1,1,1,6,9]}^4$ model. The plot represents a total of 28,683 vacua obtained by reducing  the EFT  along the monodromy direction $v^i =(1,1)$.  We indicated in red  vacua with large  ($>5\%$) instanton corrections to the K\"ahler metric and $m_{3/2}$, and in blue  (17,054 solutions) those with corrections  $<5\%$. } 
    \label{fig:scatterPlot}
\end{figure}

The procedure described above allows us to further reduce the complex structure sector to a single field. Consider for definiteness the truncation ansatz defined by the monodromy direction $v^i=(1,1)$. The resulting effective prepotential \eqref{eq:effPrep} is given by the couplings $\kappa_{vvv}=21$ and $\kappa_v= \ft{23}{4}$. In order to construct the vacua ensemble, we first generated the collection of all flux tuples $\{f^B_0, h^B_0, \hat f_{A,B}, \hat h_{A,B}\}$ with entries in the interval $[-15,15]$ satisfying the tadpole constraint ($\sim 2\cdot 10^7$ in total). For each of these, 
we numerically solved the $F$-flatness conditions \eqref{eq:noScale} of the reduced model given by \eqref{eq:effPrep} and \eqref{eq:effFlux},  employing the  software \texttt{Paramotopy} \cite{bates2018paramotopy,SW96,NSSP}. The resulting set of 17,054 solutions  is displayed in  figure~\ref{fig:scatterPlot} (blue dots), which shows the distribution of  vacua on a fundamental domain of   $\{\tau, \hat z\}$.  This ensemble includes only solutions at the weak string coupling/LCS regime, i.e., where $g_s <1$ and  with small instanton corrections to the prepotential  (using the similar criteria to \cite{Blanco-Pillado:2020wjn}). It is interesting to note that for this particular branch of vacua, the continuous flux approximation \eqref{eq:Nvac} predicts $\cN|_{v^i=(1,1)} \sim 10^5$, which represents a very small fraction of the total number of vacua, $\cN_{\text{total}}\sim 10^{13}$. This result shows the efficiency of our method, which led us to find a significant fraction of this branch of  solutions, regardless of them arising with low frequency in the landscape.

 As we detailed before,  the truncation ansatz $z^i = v^i \hat z$ together with   \eqref{eq:constrainedFluxes} allows us to lift  each of these solutions to a  vacuum of the complete $\mathbb{WP}_{[1,1,1,6,9]}^4$ model.
After the lift, we computed the scalar mass spectrum at each vacuum  for the axio-dilaton and the  $\cG$-invariant $z^i$ modes ($\lambda = 0,1,2$) by direct diagonalisation of the Hessian of the flux potential of the  $\mathbb{WP}_{[1,1,1,6,9]}^4$ model. The result perfectly matched the formula \eqref{eq:N0Spectrum} in all cases. It is important to emphasise that, at each of the obtained  solutions, equation \eqref{eq:N0Spectrum} also gives the masses of the 270 truncated complex fields which transform non-trivially under $\cG$, i.e., the modes with $\lambda=3,\ldots,272$. This is a remarkable result, given that we only used the EFT couplings for the $\cG$-invariant moduli computed in  \cite{Candelas:1994hw}.  A statistical analysis of the flux vacua obtained by this procedure  can be found in  appendix \ref{app:statistics}. 

\section{Discussion}

In this paper, we presented a method to construct ensembles of flux vacua for generic Calabi-Yau compactifications at LCS, where the masses of the axio-dilaton and complex structure moduli are given by the universal formula \eqref{eq:N0Spectrum}. This result provides full analytic control,  to leading order in $\alpha'$ and $g_s$,  over the masses of those  fields,  and therefore the  vacua we consider are an excellent stepping-stone towards  the complete stabilisation of the compactification, i.e., including  the K\"ahler moduli.  
Interestingly, up to an overall scale, the masses given in \eqref{eq:N0Spectrum} are completely determined by the vacuum values of the complex structure fields.
  As a consequence, knowing the magnitude of the $\alpha'$ and nonperturbative corrections  which generate the K\"ahler moduli  potential, it is possible to guarantee the stability of 
the axio-dilaton and all the complex structure fields by restricting  the search of vacua to appropriate  regions of moduli space.
In particular,  the spectrum \eqref{eq:N0Spectrum} involves a single asymptotically massless mode in the neighbourhood of the LCS point, $\mu_{-1}^2|_{\xi\to 0}=0$ \cite{Blanco-Pillado:2020wjn}, with
 all the remaining masses  being at least of the order of the gravitino mass $m_{3/2}$. In other words, in the LCS limit  there is only one potentially dangerous mode which might threaten the stability of the compactification. On the contrary, away from the LCS point the mass of the lightest mode in \eqref{eq:N0Spectrum} becomes of the order of $m_{3/2}$,  and thus 
 as  long as the perturbative and nonperturbative contributions  to the EFT are under control, 
  the final vacuum with the K\"ahler moduli fixed will not develop an instability.
Nevertheless, it is expected that such contributions will induce small corrections in the spectrum \eqref{eq:N0Spectrum} which, in particular, will lift the degeneracy of the modes $\lambda =2,\ldots, h^{2,1}_-$.     
 
As a final remark, note that the class of vacua we discussed is only appropriate for the construction of LVS solutions, but not for the KKLT scenario. For the solutions presented here, the flux superpotential satisfies $W_0 \equiv \cV m_{3/2} \ge 1/\sqrt{\pi}$ (see eq.~\eqref{eq:gravitino}), while the KKLT vacua require $W_0$ to be exponentially small.
Therefore, a logical future direction would be to consider the stabilisation of the K\"alher moduli at the class of vacua presented here within the LVS framework.
Another interesting continuation of this work, would be to study other truncation schemes compatible with the more general vacua discussed in \cite{Blanco-Pillado:2020wjn}, where the spectrum can also be explicitly computed, and $W_0$ could be arbitrarily small.\\

{\bf Acknowledgements}\textemdash \, 
This work is supported 
by the Spanish Ministry MCIU/AEI/FEDER grant (PGC2018-094626-B-C21), the Basque Government grant (IT-979-16) 
and the Basque Foundation for Science (IKERBASQUE). KS is supported by the Czech science foundation 
GA\v CR grant (19-01850S). MAU is also supported by the University of the Basque Country grant (PIF17/74). 
For the numerical work we  used the computing infrastructure of the 
ARINA cluster at the University of the Basque Country (UPV/EHU).

\appendix

\section{$F-$flatness condition for the truncated moduli} 
\label{app:Fcondition}
The proof of equation  \eqref{eq:truncation}  relies on the following property satisfied by the couplings of the prepotential \eqref{eq:F} in the LCS regime (neglecting instanton corrections)
\be
\kappa_{ijk}\, w^i \Im(z^j) \Im( z^k)  = 0,
\label{eq:yukawaOrthogonal}
\ee
where  $w^i$ is any holomorphic vector orthogonal to $\Im(z^i)$ with respect to the moduli space metric. This result was derived in \cite{Blanco-Pillado:2020wjn} (see eqs. (4.2) and (4.5) there), however as the conventions used here are slightly different to those of \cite{Blanco-Pillado:2020wjn}, for completeness we will briefly outline the proof in this appendix. We will assume the Euler number of the Calabi-Yau  to be non-vanishing, but it is straightforward to extend the argument to  the case $\chi(M_3)=0$, where  $\xi=\Im \kappa_0=0$.

 The K\"ahler potential  for the complex structure sector derived from \eqref{eq:KahlerPotential} and \eqref{eq:F} reads 
\be
K_{cs} = - \log \left(\ft{4}{3}  \kappa_{ijk} \Im(z^i) \Im(z^j) \Im(z^k) -2 \Im (\kappa_0) \right),
\ee
and the corresponding moduli space metric  on this  sector $K_{i \bar j} \equiv  \pd_{i} \pd_{\bar j} K_{cs}$  is 
\bea
K_{i \bar j} &=& -2 \rme^{K^{cs}}\kappa_{ijk} \Im (z^k) +  \\
&& 4 \rme^{2 K^{cs}} \kappa_{ilm} \kappa_{jnp}\Im (z^l) \Im(z^m) \Im(z^n) \Im(z^p) \nonumber.
\eea
For the instanton contributions to the prepotential \eqref{eq:F} to be suppressed, i.e., in the LCS regime, the field configuration must be such that  $\Im(z^i)$ lies in the K\"ahler cone of the mirror Calabi-Yau  (see \cite{Hosono:1994ax}), what in particular implies that  $\kappa_{ijk} \Im(z^i) \Im(z^j) \Im(z^k)$ must be non-vanishing and positive . 

Let us consider now the product $K_{i\bar j}  u^i  \Im(z^j)$ where $u^i$ is a generic holomorphic vector  not necessarily orthogonal to $\Im(z^i)$. Using that  the tensor $\kappa_{ijk}$ is totally symmetric, and the definition  $\xi \equiv \frac{-2 \rme^{K_{cs}} \Im \kappa_0}{1+ 2 \rme^{K_{cs}} \Im \kappa_0}$\footnote{This definition reduces  to the one in the main text when  $z^i= \hat z v^i$. }, this product can be expressed as
\be
K_{i\bar j}u^i \Im(z^j)  = -\frac{\xi (1 -2 \xi)}{(1+\xi)^2 \Im \kappa_0} \kappa_{ijk} u^i \Im(z^j) \Im(z^k). 
\label{eq:product}
\ee
Setting $u^i= \Im(z^i)$ in the previous equation we obtain
\be
K_{i\bar j} \Im(z^i) \Im(z^i) =\frac{3 (1-2 \xi)}{4(1+ \xi)^2},
\ee
implying that at field configurations where $\xi=1/2$ the moduli space metric becomes degenerate and thus  the  EFT is not well defined. Moreover, the case $|\xi|\to \infty$ corresponds to configurations outside the LCS regime where $\kappa_{ijk} \Im(z^i) \Im(z^j) \Im(z^k)=0$, and thus the EFT defined by the polynomial prepotential \eqref{eq:F} cannot be trusted\footnote{Actually, all field configurations with $\xi<-1$  or $\xi >1/2$ are unphysical since the field space metric always has at least one negative eigenvalue there \cite{Blanco-Pillado:2020wjn}.}. Finally, to prove equation \eqref{eq:yukawaOrthogonal} we substitute  $u^i=w^i$ in \eqref{eq:product}, with  $w^i$ orthogonal to $\Im(z^i)$, leading to 
\be
-\frac{\xi (1 -2 \xi)}{(1+\xi)^2 \Im \kappa_0} \kappa_{ijk} w^i \Im(z^j) \Im(z^k) =0,
\ee
which can only be vanishing at  physical configurations away from the LCS point ($\xi=0$) provided \eqref{eq:yukawaOrthogonal} is satisfied.

\section{Vacua statistics for the    $\mathbb{WP}_{[1,1,1,6,9]}^4$  hypersurface} 
\label{app:statistics}

In this appendix, we will discuss the statistical properties of the class of solutions presented in the main body. In particular, we will analyse the  distribution of these vacua on the reduced moduli space $\{\tau, \hat z\}$, and the probability density functions for the scalar masses in \eqref{eq:N0Spectrum}. 
For this purpose, following the method presented before, we numerically constructed an ensemble of flux vacua on an orientifold of the Calabi-Yau hypersurface $\mathbb{WP}_{[1,1,1,6,9]}^4$, and extracted the corresponding probability distributions by the direct examination  of this set of solutions.  As shown below, these numerical results are in good agreement with the analytical probability distributions derived  
in \cite{Blanco-Pillado:2020wjn}, which describe the statistics of  compactifications  with an effectively one-dimensional complex structure sector (see also \cite{Denef:2004ze}). It is important to emphasize that our analytical description of the probability distributions is independent of the choice of Calabi-Yau and the truncation ansatz. Therefore, the statistical features observed here for the $\mathbb{WP}_{[1,1,1,6,9]}^4$ ensemble are expected to be present in generic compactifications as well.

In order to have a sufficiently large sample of vacua for the statistical analysis we considered the effective reduction of the complex structure sector along the monodromy directions 
$v^i = \{(1,1),(1,2),(1,3)\}$, and we combined in a single ensemble the solutions to the  $F$-flatness conditions \eqref{eq:noScale} found for each of the three cases. Other families could also have been considered; however, the study of any of them is very computationally demanding and, due to the universal features of these vacua, we do not expect to gain any new information from studying a different family.

Furthermore, to relax the tadpole constraint on the fluxes, we considered the setting adopted in \cite{Denef:2004dm}, where  the type-IIB compactification  on $\mathbb{WP}_{[1,1,1,6,9]}^4$ was regarded as the orientifold limit of  $F$-theory on an elliptically fibered Calabi-Yau fourfold, $M_4$. In the $F$-theory framework, the maximum allowed $D3$ charge induced by the fluxes is determined by the Euler number of the fourfold, leading in the present case  to\footnote{The caveat on this approach is that it introduces additional  $D7$-brane moduli fields. For simplicity, here we will ignore  those additional moduli, and we refer the reader to \cite{Lust:2005bd,Collinucci:2008pf,Alim:2009bx,Honma:2017uzn,Honma:2019gzp,Bena:2020xrh} for discussions on their stabilisation.}  $Q_{D3} \le \chi(M_4)/24 = 273$  \cite{Denef:2004dm}.

For each choice of $v^i$, we proceeded in a similar way as described in the main body of the paper. First, we generated a collection of $10^7$ flux tuples $\{f^B_0, h^B_0, \hat f_{A,B}, \hat h_{A,B}\}$ drawn from an uniform distribution with support in $\[-25,25\]$,  and subject to  the tadpole constraint   $Q_{D3}\le 273$. Then  we searched for solutions to the corresponding $F$-flatness equations \eqref{eq:noScale} with the aid of the  software  \texttt{Paramotopy}.  The resulting ensemble contains  $206{,}479$ vacua  in the weak string-coupling regime, i.e., with $(\Im \tau)^{-1} = g_s <1$, out of which $95{,}626$ are in the LCS regime. Here we defined the LCS regime by the condition that the leading instanton contributions to the prepotential  \eqref{eq:F}, given by  (see \cite{Candelas:1994hw})
\be
\cF_{\text{inst}} = -\frac{135}{2 \pi^3} \rmi\rme^{\rmi 2 \pi z^1} -\frac{3}{8 \pi^3} \rmi \rme^{\rmi 2 \pi z^2} + \ldots \, ,
\label{eq:instantonF}
\ee
induce small relative corrections  ($<5\%$) to moduli space geometry (i.e., to the field space metric and the canonically normalised  couplings $\kappa_{ijk}$) and to the gravitino mass  $m_{3/2}$   \cite{Blanco-Pillado:2020wjn}.  
It is important to mention  that  our  definition of the LCS regime is more restrictive than just requiring $\cF_{\text{inst}}$  to be small (in absolute value) with respect to the perturbative part of the prepotential \eqref{eq:F} (see, e.g., \cite{MartinezPedrera:2012rs}). Indeed, the moduli space metric becomes degenerate far from  the LCS point
  ($\xi\to-1$ for $\chi(M_3)>0$ and $\xi\to 1/2$ for $\chi(M_3)<0$) and thus, in that regime, the metric eigenvalues are small and very sensitive to the instanton corrections, even for small  ratios $|\cF_{\text{inst}}/\cF| \sim 0.01$.

The method used here for
 avoiding duplicities in the counting of vacua is essentially  the  same as the one used in \cite{Blanco-Pillado:2020wjn} (see also \cite{DeWolfe:2004ns}). However, the case at hand requires certain specific considerations related to 
 the truncation of the moduli space so, for completeness,  we will briefly summarise this method  in the next section.

\subsection{Redundancies and solution duplicates}

The description of the EFT presented in the main text has two inherent redundancies, namely those associated to the choice of holonomy basis \eqref{eq:symplecticTrans}, and the well known $\mathrm{SL}(2,\mathbb{Z})$ modular transformations acting on $\tau$.  Those vacua which can be related to each other by these gauge transformations should be regarded as  physically equivalent, and thus when constructing the ensemble one must ensure that each distinct solution is only counted once. 

Regarding the choice of holonomy basis,  the coefficients $\kappa_{ij}$, $\kappa_i$ and $\kappa_0$ are only defined modulo integers, with different representatives associated   
to different choices of this basis. Therefore, by selecting a particular expression for  the prepotential the symplectic gauge is partially fixed, with the residual gauge given by the monodromy transformations around the LCS point, i.e.,  $z^i \to z^i +\delta^i_p$ with $p\in 1,\ldots, h^{2,1}_-$.  As a result of imposing the 
truncation ansatz  $z^i = \hat z v^i$, the gauge freedom is further reduced, leaving as the only source of gauge redundancy  the monodromy transformations $z^i \to z^i + v^i$, which  amounts to the shift
\be
 \hat z \to \hat z +1
\label{eq:redMonodromyZ}
 \ee
 on the field surviving the truncation. The corresponding symplectic transformation  $\cS_{(v)}\in \mathrm{Sp}(2 h^{2,1}_-+2,\mathbb{Z})$, acts on the period vector as (see \cite{Candelas:1993dm})
\be
\Pi(z^i + v^i) = \cS_{(v)} \cdot \Pi(z^i), \quad \text{with}\quad \cS_{(v)}  \equiv \begin{pmatrix}
A & B \\
0 & C
\end{pmatrix}.
\label{eq:residualS}
\ee
The matrices $A$ and $B$ are given by 
\be
 A = \begin{pmatrix}
\mathbb{1} & -v^i\\
0&\mathbb{1}
\end{pmatrix},
\quad B = \begin{pmatrix}
2 \kappa_v + \ft16 \kappa_{vvv} & -\kappa_{jv}+\ft12 \kappa_{jvv}\\
-\kappa_{iv} -\ft12 \kappa_{ivv}  & -\kappa_{ijv}  
\end{pmatrix},
\label{eq:monodromyMatrix}
\ee
and  $C = (A^T)^{-1}$.  Note that the condition \eqref{eq:couplingConstraint} is necessary for $\cS_{(v)}$ to have integer entries, which also requires the additional constraint $2 \kappa_v + \ft16 \kappa_{vvv} =0 \mod \mathbb{Z}$  \cite{Mayr:2000as}.

Finally, to obtain the action of the residual monodromy transformation \eqref{eq:residualS} on the fluxes of the reduced theory, we just need to impose the ansatz  
\eqref{eq:constrainedFluxes} together with   \eqref{eq:symplecticTrans}.  We find the transformation rules
\bea
\hat N_A&\to& N_A, \nonumber\\
\hat  N^B &\to&  \hat N^B - q^{-1}  \hat N_A,  \nonumber\\
N^B_0 &\to& N^B_0  +  \kappa_{vvv} \left( \hat N_A - q \hat N^B \right),
\label{eq:monodromyRed}
\eea
The condition $N_A^0=0$ is preserved.  

In addition to these transformations, one must  also take into account the modular  transformations $SL(2,\mathbb{Z})$,  which act   on the axio-dilaton and the fluxes as
\be
\tau \to \frac{a \tau + b}{c \tau +d}, \qquad \begin{pmatrix}f\\ h\end{pmatrix} \to \begin{pmatrix}a &b\\
c& d\end{pmatrix}\cdot \begin{pmatrix}f\\ h\end{pmatrix},
\label{eq:modular}
\ee
with $a,b,c,d\in\mathbb{Z}$ and $ad-bc=1$. 

In order to eliminate equivalent solutions related by the transformations  \eqref{eq:residualS} and \eqref{eq:modular}, all the vacua in the ensemble were transported to a fundamental domain defined by $\Re \tau \in [-1/2, 1/2)$, $|\tau|>1$, and $\Re \hat z \in [-1/2,1/2)$ using \eqref{eq:redMonodromyZ}, \eqref{eq:monodromyRed} and \eqref{eq:modular}. Once in the fundamental domain duplicate solutions are easily identified and discarded, as they correspond to those with  the same configuration for the fields and the fluxes.  The result of this procedure for the ensemble of vacua discussed in the main text was displayed in  figure~\eqref{fig:scatterPlot}. The corresponding distribution of vacua on the fundamental domain for the ensemble analysed in this appendix  shows no significant differences with respect  to  figure~\eqref{fig:scatterPlot}, and thus we have not displayed it here.

\subsection{Analytic formulae and numerical results}

We now turn to the analysis of the statistical properties of the ensemble.  As shown in \cite{Blanco-Pillado:2020wjn},  for compactifications with an effectively one-dimensional complex structure sector and  a large $D3$-charge tadpole, $Q_{D3}|_{\text{max}}\gg1$,  the statistics of the flux ensemble can be accurately described using the continuous flux  approximation of \cite{Denef:2004ze}. This approximation consists in neglecting the quantization of the fluxes, which are then treated as continuous random variables with a uniform distribution, only subject to the  tadpole constraint $f^T\cdot \Sigma \cdot h \le Q_{D3}|_{\text{max}}$. Using this simplification as the starting point, it is possible to derive
  the following expression for the distribution of vacua in the reduced complex structure space   \cite{Blanco-Pillado:2020wjn}  
  \be
\rho(\xi) d\xi = \cN \cdot \frac{(1+\xi)}{(2-\xi)^2\xi^{2/3}} d\xi\,, \quad \xi \equiv  \frac{-3 \Im (\kappa_0)}{2 \kappa_{vvv} \Im(\hat z)^3},
\label{eq:xiDistNA0}
\ee
where, for convenience, we have given the  distribution of $\Im(\hat z)$ in terms of the parameter $\xi$.
In the previous expression and the following ones, $\cN$ represents a normalisation constant which should be determined for each particular distribution.  It is remarkable that this distribution is independent of the details of the Calabi-Yau orientifold, or the choice of the surviving field in the reduced theory, i.e., of  $v^i$.  As a consequence, this expression can be used to describe mixed ensembles containing vacua from different compactifications and/or obtained from different truncation ansatze. 

\begin{figure}
    \centering 
	\includegraphics[width=0.43\textwidth]{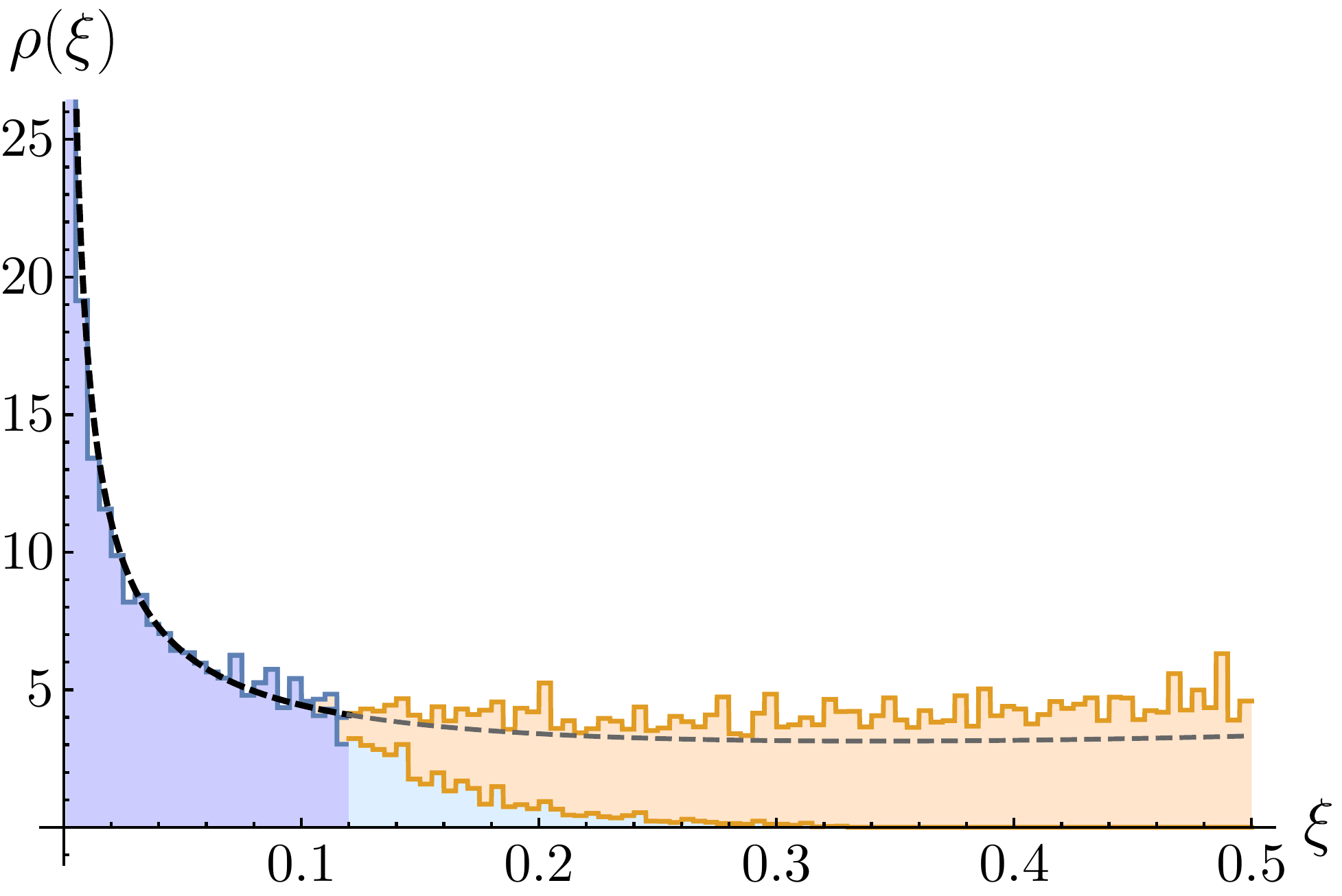} 
    \caption{Density of flux vacua in the reduced complex structure space in terms of the parameter $\xi$. The plot  shows the numerical distribution obtained directly form the ensemble of 206,479
    flux vacua. We have indicated in (dark and light) blue  the 95,626 vacua with small ($<5\%$) instanton corrections, and in orange those where the instaton contribution is large ($>5\%$).   The dashed line represents the analytic distribution  \eqref{eq:xiDistNA0} normalised in the range  $\xi \in [0.001,0.12]$ where most vacua are in the LCS regime and  the continuous flux approximation holds (dark blue).} 
    \label{fig:xiDist}
\end{figure} 

The distribution of values for the parameter $\xi$ obtained numerically for our ensemble of vacua, combining the cases $v^i=\{(1,1),(1,2),(1,3)\}$, is displayed in figure \ref{fig:xiDist}. The figure shows a stacked histogram with   the 95,626 vacua at the LCS  regime indicated in  (light and dark) blue, 
 and in orange  those solutions with a large contribution from instantons ($>5\%$). Since the formula \eqref{eq:xiDistNA0} was obtained while completely ignoring the contribution from instantons, it is expected to work only in the regime of $\xi$ space where few  vacua, or none, are discarded due to having large corrections (that is, for $\xi \lesssim 0.12$). Furthermore, due to the limitations of our numerical method, the flux integers in the ensemble range only in the interval $\[-25,25\]$, leading to an artificial bound to how close the vacua in the ensemble  can be to the LCS point, $\xi\gtrsim 0.001$ \cite{Blanco-Pillado:2020wjn}. As a consequence,  the distribution  \eqref{eq:xiDistNA0} is expected to describe correctly the statistics of
 vacua in the range $\xi \in [0.001,0.12]$,  which we have  indicated in  figure~\ref{fig:xiDist} in dark blue.

 The distribution \eqref{eq:xiDistNA0}, normalised in  its range of validity, is also indicated in the figure with a dashed line and, as it can be observed,  it provides a very good description for the density  of flux vacua. It is also interesting to note that, despite the divergence of the distribution \eqref{eq:xiDistNA0} at $\xi=0$, this function is normalizable in $\xi\in [0,1/2)$, and thus it predicts a finite number  of vacua in any  neigbourhood of the LCS point.

\begin{figure}
	\centering 
	\includegraphics[width=0.43\textwidth]{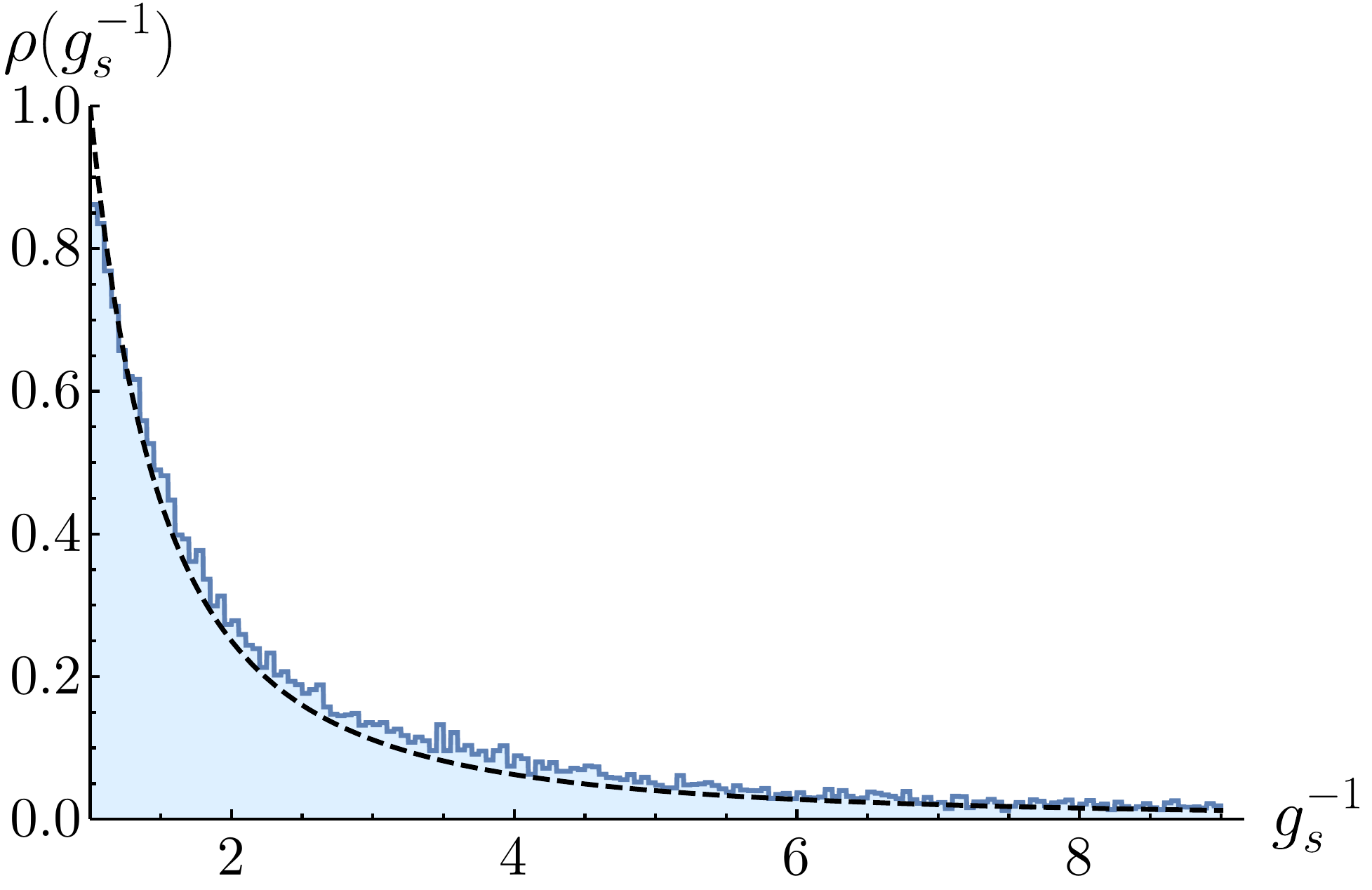} 
	\caption{Distribution of the string coupling $g_s$, in terms of $g_s^{-1}=\Im \tau$. The histogram represents the normalised distribution of the data points lying in the interval $\xi \in [0.001,0.12]$, while the dashed curve is the expected result from the continuous flux approximation} 
	\label{fig:gsDist}
\end{figure}

  Regarding the axio-dilaton, it can also be shown that, according to the continuous flux approximation,  the  string coupling constant $g_s = (\Im \tau)^{-1}$ has a uniform probability distribution in this class of vacua or, equivalently, the probability density function for the imaginary part of $\tau$ is of the form  $\rho(\Im \tau) \propto (\Im \tau)^{-2}$. This is also consistent with the distribution which we obtained numerically, as it can be seen in figure~\ref{fig:gsDist}.

In order to write the expression for the mass distributions, it is convenient to define the following functions of the parameter $\xi$:
\be
 \tilde m_\lambda(\xi) =
\left\{
\begin{array}{l c l }
\sqrt{(1-2 \xi)/3}\; \hat m(\xi)&\qquad & \lambda=0,\\
 \frac{\sqrt{(1-2 \xi)}}{\sqrt{3}\hat m(\xi)} &\qquad& \lambda=1,\\
 \frac{1+\xi}{3} &\qquad& \lambda=2,\ldots, h^{2,1}_-,
\end{array}
\right.\,
\label{eq:gralFermionSpectrum32}
\ee
which  give the fermion masses    at no-scale vacua, $m_\lambda$,  normalised by the gravitino mass $\tilde m_\lambda \equiv m_\lambda/m_{3/2}$ \cite{Blanco-Pillado:2020wjn}. Then, combining the previous expressions with \eqref{eq:xiDistNA0} and using that the functions $\tilde m_\lambda(\xi)$ are monotonic, it is immediate to obtain $h^{2,1}_-+1$ separate probability distributions, one for each of the rescaled  fermion masses 
\be
\rho^f_\lambda(\tilde m_\lambda) d\tilde m_\lambda = \cN \cdot \frac{(1+\xi)}{(2-\xi)^2\xi^{2/3} \, (d\tilde m_\lambda(\xi)/d\xi)}\Big|_{\xi(\tilde m_\lambda)} d \tilde m_\lambda\,,
\label{eq:fermionSpectrumNA0}
\ee
where  $\lambda = 0, \ldots, h^{2,1}_-$. Finally, from  the relation 
\be
\mu^2_{\pm\lambda} = (m_{3/2}^2 \pm  m_{\lambda})^2
\label{eq:scalarMasses}
\ee
 between the  scalar and  fermion masses  \cite{Sousa:2014qza}, we can obtain $h^{2,1}_-+1$ separate probability distributions, one  for each pair of normalised scalar masses $\tilde \mu_{\pm \lambda}^2 \equiv \mu_{\pm\lambda}^2 /m_{3/2}^2$ 
\be
\rho_\lambda^s(\tilde \mu^2_\lambda)  d\tilde \mu^2_\lambda= \cN\cdot \tilde \mu^{-1}_\lambda \, \left[ \rho^f_\lambda(1+\tilde \mu_\lambda) +\rho^f_\lambda(|1-\tilde \mu_\lambda|) \right] d\tilde \mu^2_\lambda.
\label{eq:scalar_mass_spectrum}
\ee

\begin{figure}[th]
	\centering
	\subfloat[]{
		\includegraphics[width=0.43\textwidth]{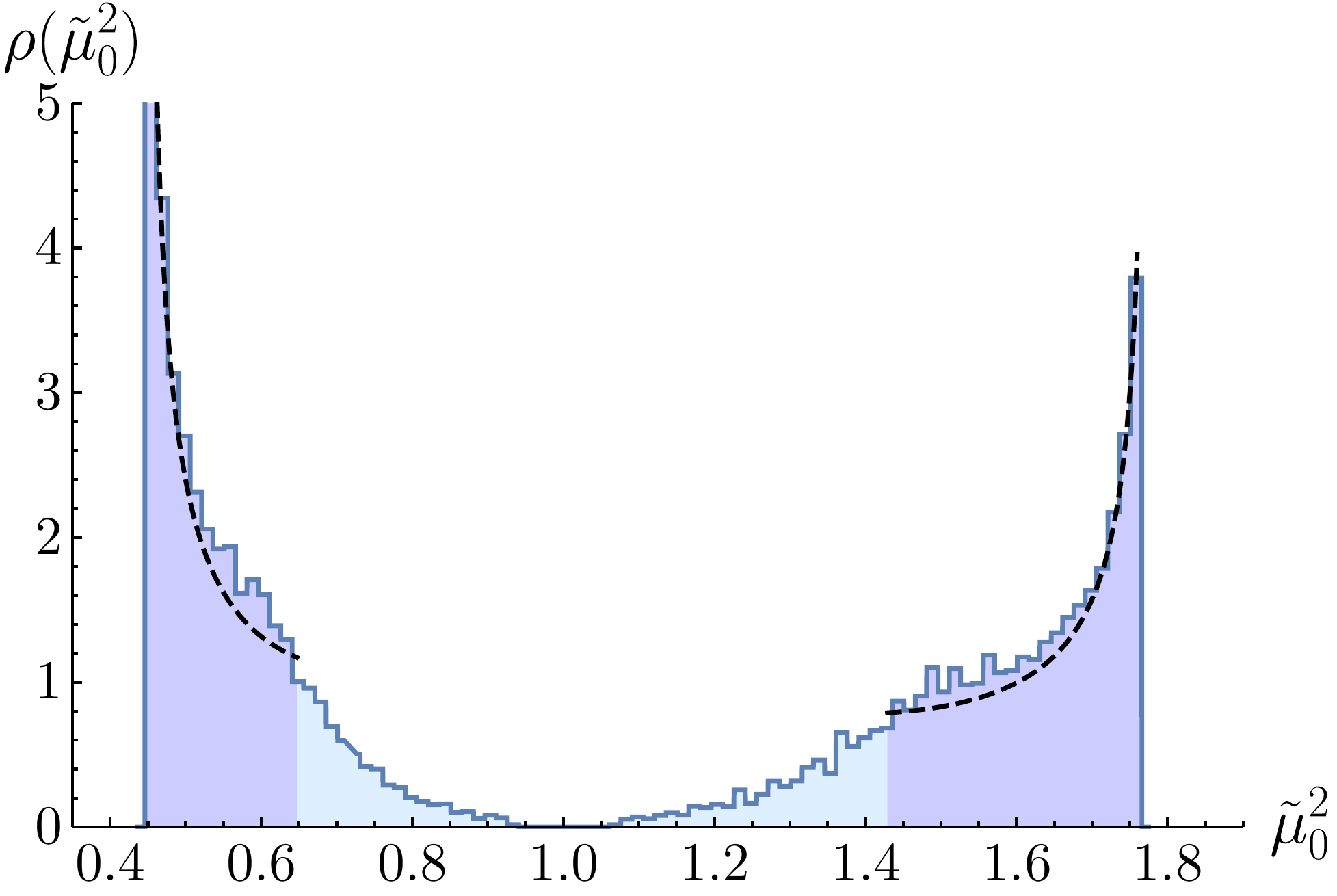}
	}
	\hfill  
	\subfloat[]{
		\includegraphics[width=0.43\textwidth]{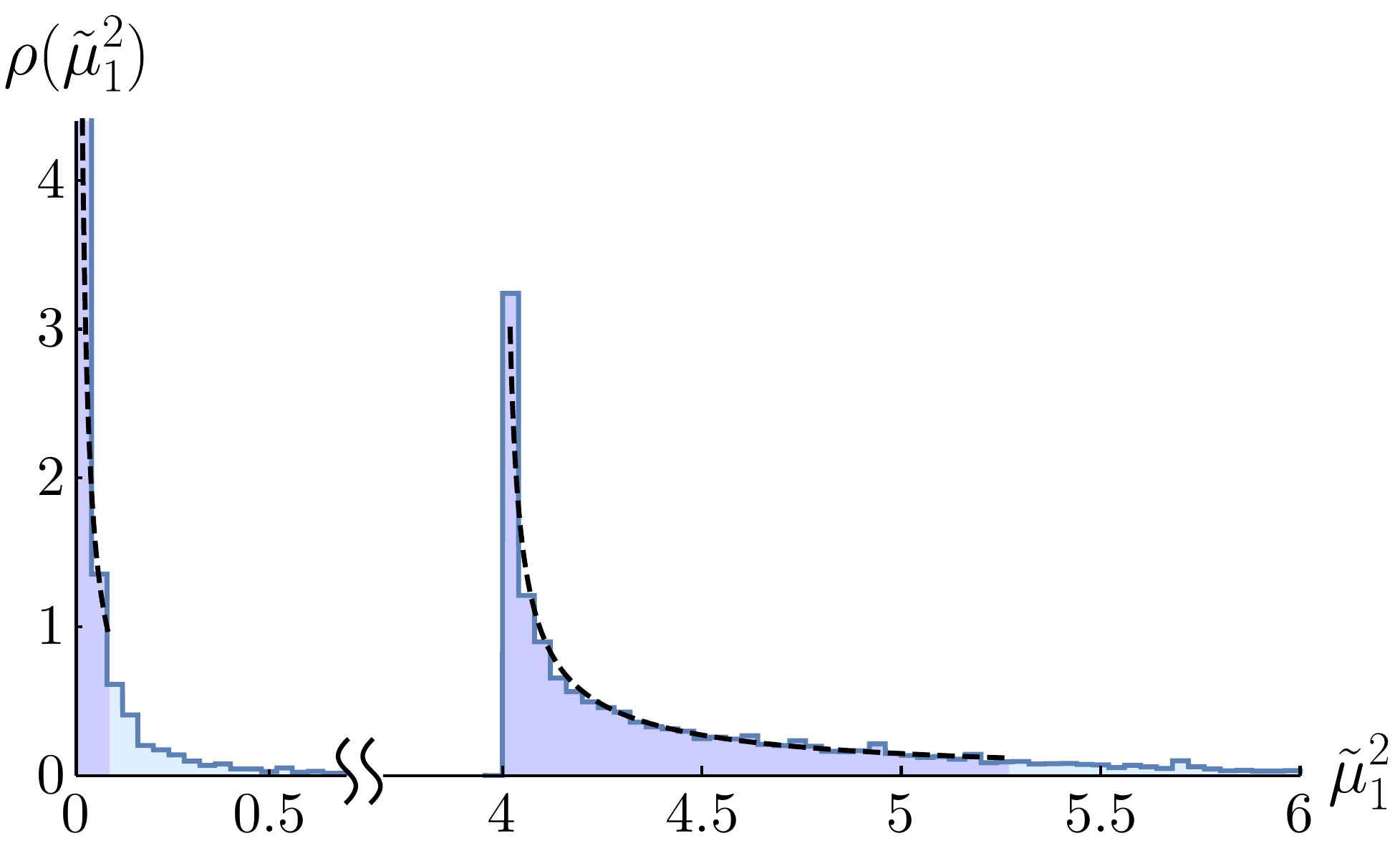}
	}
	\hfill  
	\subfloat[]{
		\includegraphics[width=0.43\textwidth]{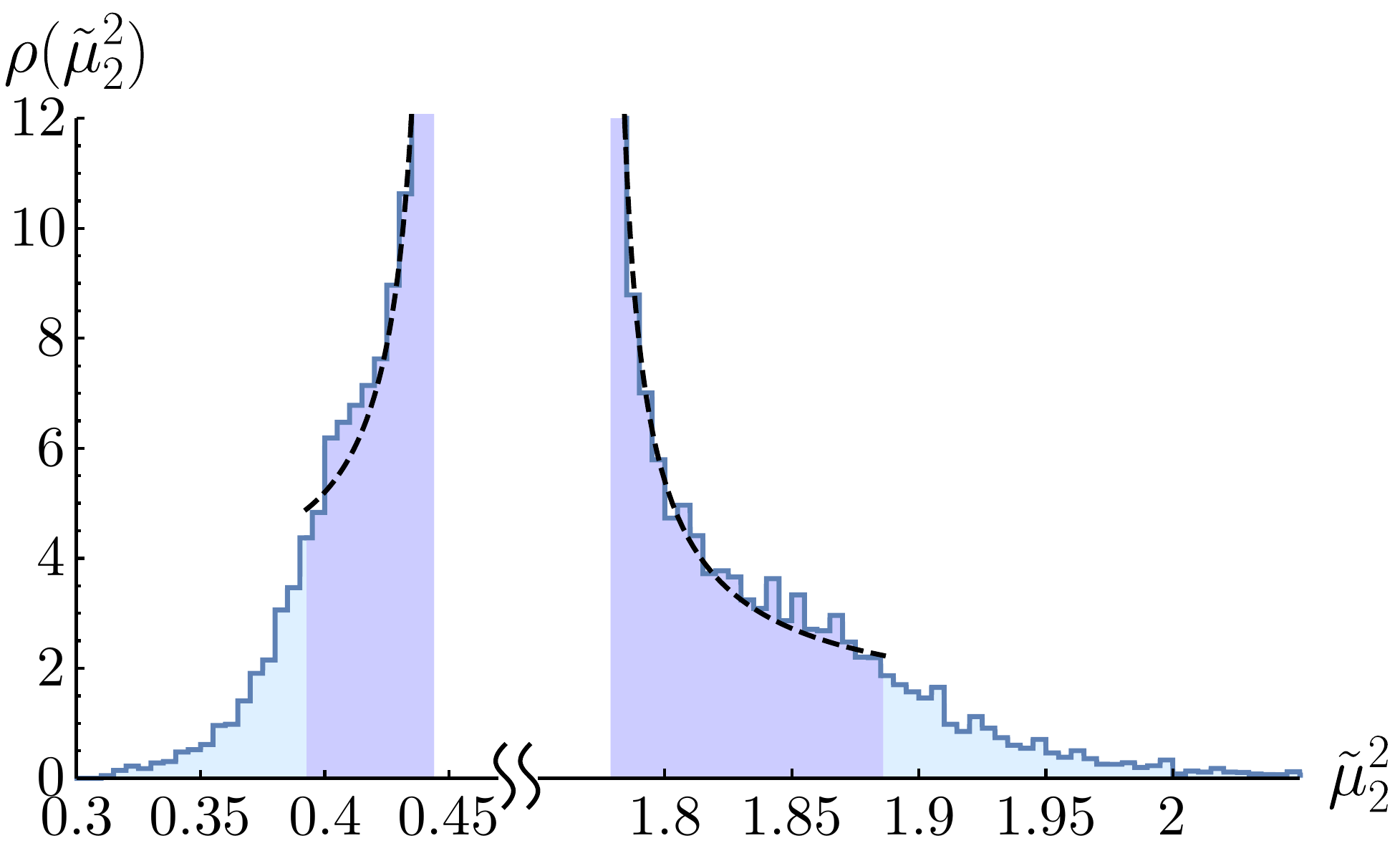}
	}
    \caption{Numerical distributions for the normalised scalar masses $\tilde \mu^{2}_{\pm\lambda}=\mu^{2}_{\pm\lambda}/m_{3/2}^2$ of the $\cG-$invariant modes, $\lambda=\lbrace 0,1,2 \rbrace$ in the ensemble of 95,626 vacua at LCS. In each figure, the dashed line represents the analytic formula \eqref{eq:scalar_mass_spectrum} for each value of $\lambda$, normalised in the same range $\xi \in [0.001,0.12]$.  
    The darker regions represent the solutions for which the continuous flux approximation applies.} 
    \label{fig:massDist}
\end{figure} 
In order to generate the numerical mass distributions for our ensemble of vacua, at each solution to \eqref{eq:noScale} we diagonalised the Hessian of the scalar potential induced by the fluxes, i.e., the potential in  the theory defined by \eqref{eq:KahlerPotential}, \eqref{eq:F}, and \eqref{eq:EFT} with the couplings \eqref{eq:couplings}, which describe the $\cG$-invariant sector of the moduli space in the $\mathbb{WP}_{[1,1,1,6,9]}^4$ model.  
In all cases, the resulting masses for the three $\cG$-invariant modes (including the axio-dilaton) were in agreement with equation~\eqref{eq:N0Spectrum} with $\lambda=0,1,2$. The numerical distributions for the scalar $\mu^2_{\pm0}$, $\mu^2_{\pm1}$ and $\mu^2_{\pm2}$  are displayed  in figure \ref{fig:massDist}, along with the theoretical distribution \eqref{eq:scalar_mass_spectrum}  normalised in the range $\xi \in [0.001,0.12]$. As expected from the analysis of the distribution $\rho(\xi)$ \eqref{eq:xiDistNA0}, in figure~\ref{fig:massDist} we can see that the theoretical probability densities for the masses are in good agreement with the obtained numerical results. The most significant feature of these plots is that the density distribution for $\mu_{\pm1}^2$ is peaked around zero, indicating that a large fraction of vacua involve a  light field in the spectrum. This can be understood recalling that, on the one hand,  vacua with $N_A^0=0$ (as those discussed here) can be found parametrically close to the LCS point \cite{Marsh:2015zoa,Junghans:2018gdb,Grimm:2019ixq}, and thus a large fraction is expected to be found near $\xi=0$ (see figure \ref{fig:xiDist}). On the other hand, as we mentioned in the main text, from   \eqref{eq:N0Spectrum} it follows that the spectrum of these vacua  contains an asymptotically massless mode in the limit $\xi \to0$, which explains the peak of $\rho^s_1(\mu^2_{1})$ at $\mu^2_{1}=0$ observed in figure \ref{fig:massDist}(b). This feature is expected to be generic for the class of vacua discussed here, regardless of the choice of Calabi-Yau compactification or the truncation ansatz, as both the mass spectrum \eqref{eq:N0Spectrum} and the probability distributions \eqref{eq:scalarMasses} and \eqref{eq:scalar_mass_spectrum} are completely universal. Note also that half of the masses in the spectrum are smaller than the gravitino mass $m_{3/2}^2$.

\begin{figure}
	\centering
	\subfloat[]{
		\includegraphics[width=0.43\textwidth]{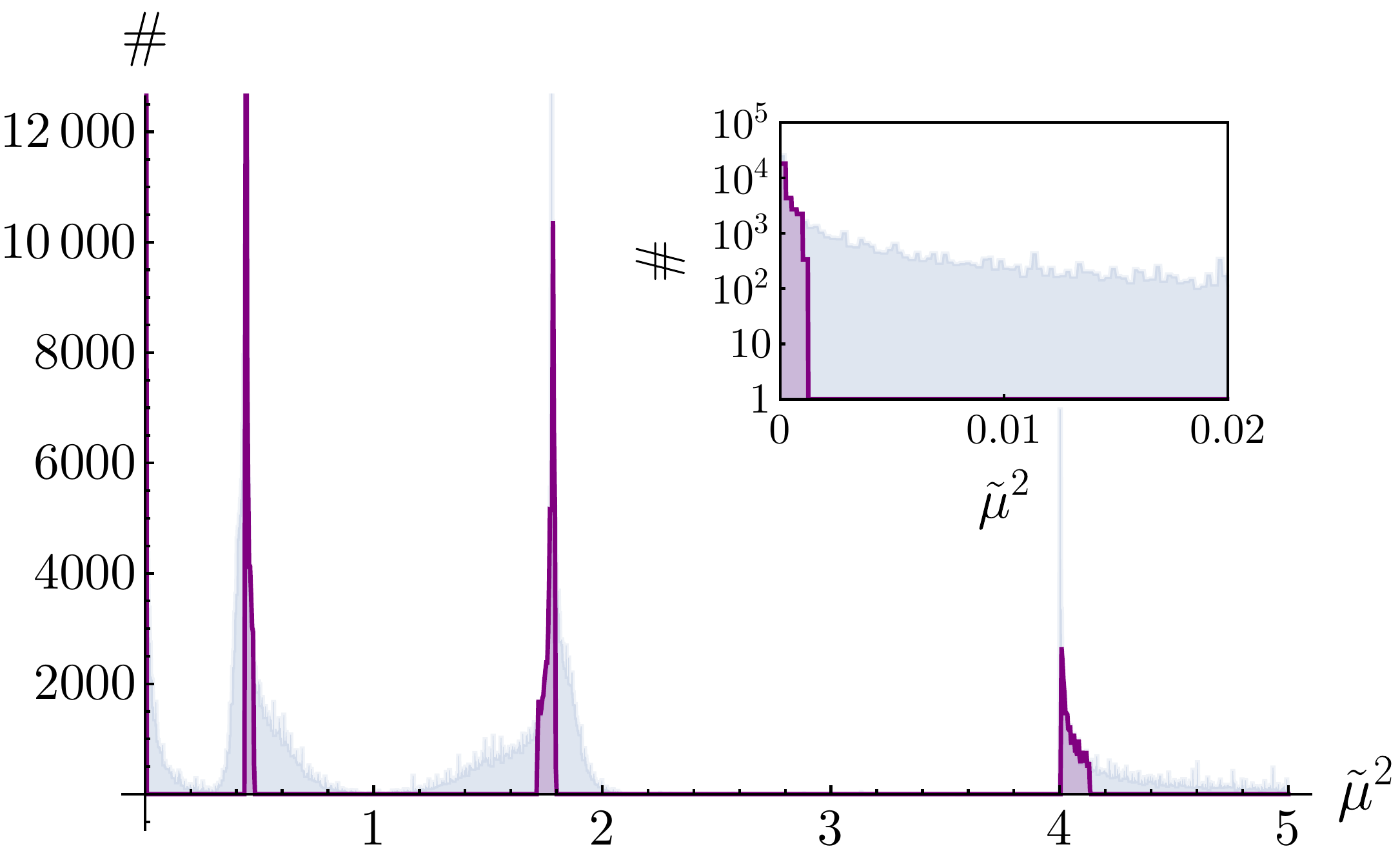}
	}
	\hfill  
	\subfloat[]{
		\includegraphics[width=0.43\textwidth]{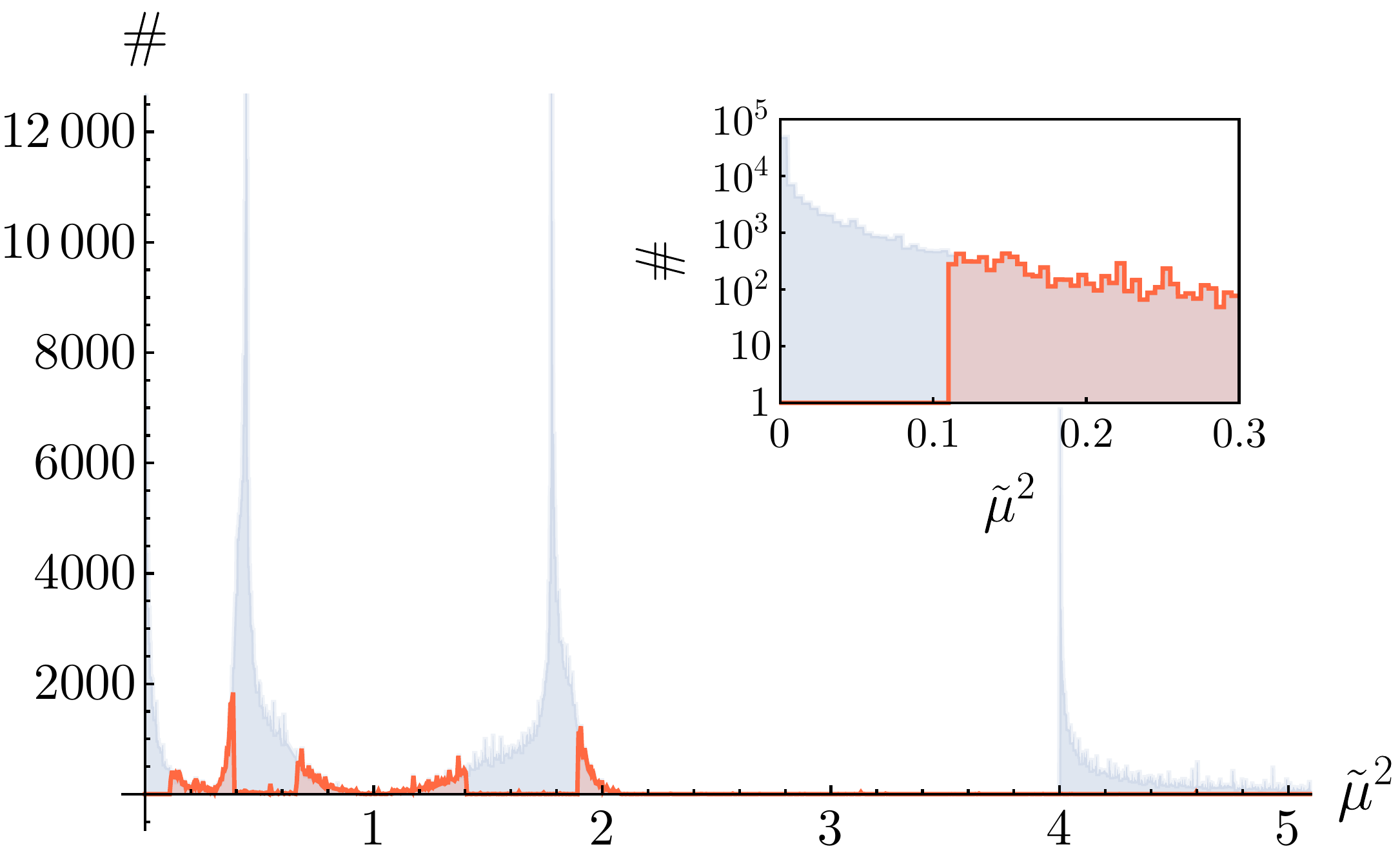}
	}
	\caption{Complete distribution of scalar masses $\tilde{\mu}^2_{\lambda}$, $\lambda=\lbrace 0,1,2 \rbrace$, in the ensemble of 95,626 vacua at LCS. In each histogram, solutions with (a) $\xi \in [0.001,0.02]$ and (b) $\xi \in [0.13,0.5)$ have been highlighted. The insets in these plots show the mass distribution near $\tilde \mu^2 \approx 0$, illustrating the presence of an asymptotically massless mode at vacua near the LCS point (a), and the absence of light modes in the spectrum when solutions near the LCS point are excluded (b).} 
	\label{fig:LCSmass}
\end{figure} 

The sharp edges of the mass sectra shown in figure \ref{fig:massDist} correspond to the cutoffs we have set on  the parameter $\xi \in[0.001,0.12]$, with  the peaks of the probability distributions corresponding to  the minimum value of $\xi$. The effect of changing the bounds of $\xi$ can be seen in figure \ref{fig:LCSmass}. The plot in figure \ref{fig:LCSmass}(a) represents the combined distribution for the masses of the three $\cG-$invariant modes, with $\xi \in [0.001,0.02]$. As is can be seen the distribution becomes very peaked, with the maxima at the values  
\be
\mu^2_{\pm \lambda}/m_{3/2}^2 = \{0,\ft49, \ft{16}{9},4 \},
\ee
which is precisely the strict LCS limit ($\xi\to 0$)  of the spectrum \eqref{eq:N0Spectrum}. This result illustrates the appearance of an asymptotically massless mode at vacua located in a small neigbourhood of the LCS point. Figure \ref{fig:LCSmass}(b) shows the combined the mass distribution for the $\cG-$invariant modes, with the parameter $\xi$ restricted to  $\xi \in [0.13,0.5)$, that is, for vacua with small instanton corrections but well separated from the LCS point. In the inset of  figure \ref{fig:LCSmass}(b) we can see that for all these vacua the mass  of the lightest mode is now bounded below $\mu^2_{-1} \gtrsim 0.11 \, m_{3/2}^2$. Then, as we mentioned in the main text, it can be seen that by excluding solutions from a neighbourhood of the LCS point we can construct an ensemble of  vacua whose  spectrum does not contain light modes,  and thus, which are good candidates for  considering the stabilisation of the K\"ahler moduli.

We note that the combined density distribution for the scalar masses given in
Fig. (\ref{fig:LCSmass}) is quite peaked around specific values. This is in contrast with the general
predictions based on the statistical modelling of a landscape where this
distribution is found using Random Matrix Theory \cite{Denef:2004cf,Marsh:2011aa,Sousa:2014qza,Achucarro:2015kja}. The
origin of this difference comes from the lack of sufficient complexity in our
ensemble of effective field theories. The reason for this is that the structure of the couplings
in our ensemble of vacua is quite rigid and does not display the sufficient random nature for the scalar potential induced by the fluxes
to be described by a multidimensional Gaussian random field.

Finally, the 270 truncated complex structure fields transforming non-trivially under the symmetry group $\cG$  have the same masses as the $\cG$-invariant modes with $\lambda=2$, namely  $\mu^{2}_{\pm \lambda} = \left(1 \pm \ft13 (1+\xi)\right)^2$ for all $\lambda=2,\ldots h^{2,1}_-$, and thus their probability distributions  coincide with the one displayed in the lower plot of figure \ref{fig:massDist}. Let us emphasize again that this  is a  rather exceptional result, as the EFT couplings for these fields  are not determined by \eqref{eq:couplings}, and thus they were  unknown a priori.

\section{Estimate of the number of vacua}
\label{app:Nvac}
In this appendix we will outline the proof of  equation \eqref{eq:Nvac}, which is an estimate for   the number of vacua in a compactification compatible with the spectrum \eqref{eq:N0Spectrum}. More specifically, we will compute the number of flux vacua in the branch determined by a vector $v^i$ in the K\"ahler cone of the mirror dual to $M_3$, and which are subject to a tadpole constraint of the form $Q_{D3}\le Q_{D3}^*$, where $Q_{D3}^*$ is the available  $D3$ charge, that is,
\be
 \cN_{v^i}(Q_{D3}\le Q_{D3}^*,g_s\le g_s^*, |\xi|\le\xi^*).
 \ee
As described in the main text, these vacua are given by the solutions to the no-scale equations \eqref{eq:noScale} in the EFT determined by the prepotential \eqref{eq:effPrep} and the flux ansatz \eqref{eq:effFlux}. The number of these solutions can counted using a generalised version of the Kac-Rice formula
\bea
\cN_{v^i}&=& \sum_{\{f,h\}} \Theta(Q_{D3} - Q_{D3}^*) \times \nonumber \\
&&\int d^{4} u\; |\det D^2 W| \, \delta^{(2)}(D_{u^A}W) \delta^{(2)}(D_{\bar u^A} \bar W), \phantom{fff}
\label{eq:NvacDer1}
\eea
where we used the shorthand $u^A = \{\tau, \hat z\}$, and we are summing over the integer flux parameters $\{f,h\}=\{f_B^0,h_B^0,\hat f_{A,B}, \hat h_{A,B}\}$. Following  \cite{Denef:2004ze} in the following we will make use of the continuous flux approximation, valid when the available $D3$ charge is large $Q_{D3}\gg1$, and which allows to replace the sums over flux integers by an integral:
\be
\sum_{{\{f,h\}}} \longrightarrow \int d^4 f \, d^4h \, \delta(f_A^0) \delta(h_A^0).   
\ee
Here the integration is done  over the  \emph{continuous variables} $\{f_{A,B}^0, h_{A,B}^0, \hat f_{A,B}, \hat h_{A,B}\}$, and the Dirac delta functions enforce the constraint $f_A^0= h_A^0=0$. Using the integral representation of the Heaviside theta, and changing from the previous real flux parameters to the complex entries of the flux vector $\hat N$ (see eq.  \eqref{eq:effFlux}),  the expression \eqref{eq:NvacDer1} can be rewritten as  \cite{Denef:2004ze,Blanco-Pillado:2020wjn}
\be
\cN_{v^i}  = \frac{1}{2 \pi \rmi} \int_C d \alpha \frac{\rme^{\alpha Q_{D3}^*}}{\alpha} \cN(\alpha).
\label{integral}
\ee
Here $C$ runs along the imaginary axis passing the zero to the right, and 
\bea
\cN(\alpha) &\equiv&\ft{1}{q^2 \kappa_{vvv}^2} \int d^4 \hat N \, d^4 \bar {\hat N} \, \delta( N_A^0) \delta(\bar { N}_A^0) \rme^{3K_d} \rme^{-\alpha Q_{D3}}  \times \nonumber \\
&& \nonumber \\ 
&& \int d^{4} u\; |\det D^2 W| \, \delta^{(2)}(D_{u^A}W) \delta^{(2)}(D_{\bar u^A} \bar W), \phantom{fff}
\label{eq:NveacDer2}
\eea
with $\rme^{K_d} = 1/(2 \Im \tau)$. After trading the complex variables $\hat N_{A,B}^I$ for the quantities
\bea
F_0 &\equiv& \sqrt{\alpha} \cV \rme^{K/2} D_0 W,\nonumber \\
F_1 &\equiv& \sqrt{\alpha} \cV \rme^{K/2} D_1 W,\nonumber \\
 Z_0 &\equiv& \sqrt{\alpha}\cV \rme^{K/2} W,  \nonumber \\
 Z_1&\equiv& \sqrt{\alpha}\cV \rme^{K/2} D_0 D_1 W,  
\eea
where the derivatives are expressed in a canonically normalised basis $\{e_0^\tau, e^{\hat z}_1\}$, the parameter $\alpha$  disappears from  equation \eqref{eq:NveacDer2}, except for an overall factor $\alpha^{-2}$ \cite{Blanco-Pillado:2020wjn}. This allows us to easily perform the integral \eqref{integral}, leading to 
\bea
\cN_{v^i} &=&\ft{(Q_{D3}^*)^3}{ 3! q^2 \kappa_{vvv}^2 } \int d^4u |g| \rme^{-K_{cs}} \times\nonumber \\&& \int (dZ_1 d\bar Z_1) \rme^{-|Z_0|^2-|Z_1|^2}|\mathcal{H}|^{1/2},
\eea
with $Z_0 =-\rmi \sqrt{3/(1-2 \xi)} \bar Z_1$.
Here we used the notation $\rme^{-K_{cs}} = -\rmi \, \Pi^\dag \cdot \Sigma\cdot \Pi$, $|g|$ is the determinant of the moduli space metric, and $|\mathcal H|$ is the determinant of the Hessian of the scalar potential along the $\{\tau, \hat z\}$ directions expressed in a canonically normalised basis. The rest of the computation can be done exactly as  explained in appendix C.4 of \cite{Blanco-Pillado:2020wjn}. We obtain the final result 
\begin{align}
\cN_{v^i} &(Q_{D3}\le Q_{D3}^*, g_s \le g_s^*, |\xi| \le \xi^*) =  \nonumber \\
& =\frac{3 \pi}{16}\frac{g_s^* |\Im \kappa_0| (Q_{D3}^*)^3}{q^2 \kappa_{vvv}^2 } \frac{ |\xi^*/\xi_0|^{1/3}}{(2-  \xi^*)}.
 \end{align}
where $\xi_0 \equiv 3|\Im \kappa_0|/2 \kappa_{vvv}$. Using that the instanton contributions to the prepotential are exponentially suppressed for $|\xi|  \le \xi^* \approx \xi_0$  (see appendix D.3 of \cite{Blanco-Pillado:2020wjn}), and that $2-\xi = \mathcal{O}(1)$, it is immediate to arrive to the estimate  in  eq.  \eqref{eq:Nvac} for the number of vacua at LCS consistent with the spectrum \eqref{eq:N0Spectrum}. Given the remarkable accuracy of the continuous flux approximation in characterising the mass spectra   (see fig. \ref{fig:massDist}), we expect the estimate in eq. \eqref{eq:Nvac} to be quite precise.

\bibliography{refs.bib}

 \newcommand{\noop}[1]{}
\begin{thebibliography}{58}%
\makeatletter
\providecommand \@ifxundefined [1]{%
 \@ifx{#1\undefined}
}%
\providecommand \@ifnum [1]{%
 \ifnum #1\expandafter \@firstoftwo
 \else \expandafter \@secondoftwo
 \fi
}%
\providecommand \@ifx [1]{%
 \ifx #1\expandafter \@firstoftwo
 \else \expandafter \@secondoftwo
 \fi
}%
\providecommand \natexlab [1]{#1}%
\providecommand \enquote  [1]{``#1''}%
\providecommand \bibnamefont  [1]{#1}%
\providecommand \bibfnamefont [1]{#1}%
\providecommand \citenamefont [1]{#1}%
\providecommand \href@noop [0]{\@secondoftwo}%
\providecommand \href [0]{\begingroup \@sanitize@url \@href}%
\providecommand \@href[1]{\@@startlink{#1}\@@href}%
\providecommand \@@href[1]{\endgroup#1\@@endlink}%
\providecommand \@sanitize@url [0]{\catcode `\\12\catcode `\$12\catcode
  `\&12\catcode `\#12\catcode `\^12\catcode `\_12\catcode `\%12\relax}%
\providecommand \@@startlink[1]{}%
\providecommand \@@endlink[0]{}%
\providecommand \url  [0]{\begingroup\@sanitize@url \@url }%
\providecommand \@url [1]{\endgroup\@href {#1}{\urlprefix }}%
\providecommand \urlprefix  [0]{URL }%
\providecommand \Eprint [0]{\href }%
\providecommand \doibase [0]{http://dx.doi.org/}%
\providecommand \selectlanguage [0]{\@gobble}%
\providecommand \bibinfo  [0]{\@secondoftwo}%
\providecommand \bibfield  [0]{\@secondoftwo}%
\providecommand \translation [1]{[#1]}%
\providecommand \BibitemOpen [0]{}%
\providecommand \bibitemStop [0]{}%
\providecommand \bibitemNoStop [0]{.\EOS\space}%
\providecommand \EOS [0]{\spacefactor3000\relax}%
\providecommand \BibitemShut  [1]{\csname bibitem#1\endcsname}%
\let\auto@bib@innerbib\@empty
\bibitem [{\citenamefont {Kachru}\ \emph {et~al.}(2003)\citenamefont {Kachru},
  \citenamefont {Kallosh}, \citenamefont {Linde},\ and\ \citenamefont
  {Trivedi}}]{Kachru:2003aw}%
  \BibitemOpen
  \bibfield  {author} {\bibinfo {author} {\bibfnamefont {S.}~\bibnamefont
  {Kachru}}, \bibinfo {author} {\bibfnamefont {R.}~\bibnamefont {Kallosh}},
  \bibinfo {author} {\bibfnamefont {A.~D.}\ \bibnamefont {Linde}}, \ and\
  \bibinfo {author} {\bibfnamefont {S.~P.}\ \bibnamefont {Trivedi}},\ }\href
  {\doibase 10.1103/PhysRevD.68.046005} {\bibfield  {journal} {\bibinfo
  {journal} {Phys.Rev.}\ }\textbf {\bibinfo {volume} {D68}},\ \bibinfo {pages}
  {046005} (\bibinfo {year} {2003})},\ \Eprint
  {http://arxiv.org/abs/hep-th/0301240} {arXiv:hep-th/0301240 [hep-th]}
  \BibitemShut {NoStop}%
\bibitem [{\citenamefont {Balasubramanian}\ \emph {et~al.}(2005)\citenamefont
  {Balasubramanian}, \citenamefont {Berglund}, \citenamefont {Conlon},\ and\
  \citenamefont {Quevedo}}]{Balasubramanian:2005zx}%
  \BibitemOpen
  \bibfield  {author} {\bibinfo {author} {\bibfnamefont {V.}~\bibnamefont
  {Balasubramanian}}, \bibinfo {author} {\bibfnamefont {P.}~\bibnamefont
  {Berglund}}, \bibinfo {author} {\bibfnamefont {J.~P.}\ \bibnamefont
  {Conlon}}, \ and\ \bibinfo {author} {\bibfnamefont {F.}~\bibnamefont
  {Quevedo}},\ }\href {\doibase 10.1088/1126-6708/2005/03/007} {\bibfield
  {journal} {\bibinfo  {journal} {JHEP}\ }\textbf {\bibinfo {volume} {0503}},\
  \bibinfo {pages} {007} (\bibinfo {year} {2005})},\ \Eprint
  {http://arxiv.org/abs/hep-th/0502058} {arXiv:hep-th/0502058 [hep-th]}
  \BibitemShut {NoStop}%
\bibitem [{\citenamefont {Conlon}\ \emph {et~al.}(2005)\citenamefont {Conlon},
  \citenamefont {Quevedo},\ and\ \citenamefont {Suruliz}}]{Conlon:2005ki}%
  \BibitemOpen
  \bibfield  {author} {\bibinfo {author} {\bibfnamefont {J.~P.}\ \bibnamefont
  {Conlon}}, \bibinfo {author} {\bibfnamefont {F.}~\bibnamefont {Quevedo}}, \
  and\ \bibinfo {author} {\bibfnamefont {K.}~\bibnamefont {Suruliz}},\ }\href
  {\doibase 10.1088/1126-6708/2005/08/007} {\bibfield  {journal} {\bibinfo
  {journal} {JHEP}\ }\textbf {\bibinfo {volume} {0508}},\ \bibinfo {pages}
  {007} (\bibinfo {year} {2005})},\ \Eprint
  {http://arxiv.org/abs/hep-th/0505076} {arXiv:hep-th/0505076 [hep-th]}
  \BibitemShut {NoStop}%
\bibitem [{\citenamefont {Gross}\ and\ \citenamefont
  {Witten}(1986)}]{Gross:1986iv}%
  \BibitemOpen
  \bibfield  {author} {\bibinfo {author} {\bibfnamefont {D.~J.}\ \bibnamefont
  {Gross}}\ and\ \bibinfo {author} {\bibfnamefont {E.}~\bibnamefont {Witten}},\
  }\href {\doibase 10.1016/0550-3213(86)90429-3} {\bibfield  {journal}
  {\bibinfo  {journal} {Nucl. Phys. B}\ }\textbf {\bibinfo {volume} {277}},\
  \bibinfo {pages} {1} (\bibinfo {year} {1986})}\BibitemShut {NoStop}%
\bibitem [{\citenamefont {Becker}\ \emph {et~al.}(2002)\citenamefont {Becker},
  \citenamefont {Becker}, \citenamefont {Haack},\ and\ \citenamefont
  {Louis}}]{Becker:2002nn}%
  \BibitemOpen
  \bibfield  {author} {\bibinfo {author} {\bibfnamefont {K.}~\bibnamefont
  {Becker}}, \bibinfo {author} {\bibfnamefont {M.}~\bibnamefont {Becker}},
  \bibinfo {author} {\bibfnamefont {M.}~\bibnamefont {Haack}}, \ and\ \bibinfo
  {author} {\bibfnamefont {J.}~\bibnamefont {Louis}},\ }\href {\doibase
  10.1088/1126-6708/2002/06/060} {\bibfield  {journal} {\bibinfo  {journal}
  {JHEP}\ }\textbf {\bibinfo {volume} {06}},\ \bibinfo {pages} {060} (\bibinfo
  {year} {2002})},\ \Eprint {http://arxiv.org/abs/hep-th/0204254}
  {arXiv:hep-th/0204254} \BibitemShut {NoStop}%
\bibitem [{\citenamefont {Sethi}(2018)}]{Sethi:2017phn}%
  \BibitemOpen
  \bibfield  {author} {\bibinfo {author} {\bibfnamefont {S.}~\bibnamefont
  {Sethi}},\ }\href {\doibase 10.1007/JHEP10(2018)022} {\bibfield  {journal}
  {\bibinfo  {journal} {JHEP}\ }\textbf {\bibinfo {volume} {10}},\ \bibinfo
  {pages} {022} (\bibinfo {year} {2018})},\ \Eprint
  {http://arxiv.org/abs/1709.03554} {arXiv:1709.03554 [hep-th]} \BibitemShut
  {NoStop}%
\bibitem [{\citenamefont {Anguelova}\ \emph {et~al.}(2010)\citenamefont
  {Anguelova}, \citenamefont {Quigley},\ and\ \citenamefont
  {Sethi}}]{Anguelova:2010ed}%
  \BibitemOpen
  \bibfield  {author} {\bibinfo {author} {\bibfnamefont {L.}~\bibnamefont
  {Anguelova}}, \bibinfo {author} {\bibfnamefont {C.}~\bibnamefont {Quigley}},
  \ and\ \bibinfo {author} {\bibfnamefont {S.}~\bibnamefont {Sethi}},\ }\href
  {\doibase 10.1007/JHEP10(2010)065} {\bibfield  {journal} {\bibinfo  {journal}
  {JHEP}\ }\textbf {\bibinfo {volume} {10}},\ \bibinfo {pages} {065} (\bibinfo
  {year} {2010})},\ \Eprint {http://arxiv.org/abs/1007.4793} {arXiv:1007.4793
  [hep-th]} \BibitemShut {NoStop}%
\bibitem [{\citenamefont {Kachru}\ and\ \citenamefont
  {Trivedi}(2019)}]{Kachru:2018aqn}%
  \BibitemOpen
  \bibfield  {author} {\bibinfo {author} {\bibfnamefont {S.}~\bibnamefont
  {Kachru}}\ and\ \bibinfo {author} {\bibfnamefont {S.~P.}\ \bibnamefont
  {Trivedi}},\ }\href {\doibase 10.1002/prop.201800086} {\bibfield  {journal}
  {\bibinfo  {journal} {Fortsch. Phys.}\ }\textbf {\bibinfo {volume} {67}},\
  \bibinfo {pages} {1800086} (\bibinfo {year} {2019})},\ \Eprint
  {http://arxiv.org/abs/1808.08971} {arXiv:1808.08971 [hep-th]} \BibitemShut
  {NoStop}%
\bibitem [{\citenamefont {Abe}\ \emph {et~al.}(2007)\citenamefont {Abe},
  \citenamefont {Higaki}, \citenamefont {Kobayashi},\ and\ \citenamefont
  {Omura}}]{Abe:2006xp}%
  \BibitemOpen
  \bibfield  {author} {\bibinfo {author} {\bibfnamefont {H.}~\bibnamefont
  {Abe}}, \bibinfo {author} {\bibfnamefont {T.}~\bibnamefont {Higaki}},
  \bibinfo {author} {\bibfnamefont {T.}~\bibnamefont {Kobayashi}}, \ and\
  \bibinfo {author} {\bibfnamefont {Y.}~\bibnamefont {Omura}},\ }\href
  {\doibase 10.1103/PhysRevD.75.025019} {\bibfield  {journal} {\bibinfo
  {journal} {Phys.Rev.}\ }\textbf {\bibinfo {volume} {D75}},\ \bibinfo {pages}
  {025019} (\bibinfo {year} {2007})},\ \Eprint
  {http://arxiv.org/abs/hep-th/0611024} {arXiv:hep-th/0611024 [hep-th]}
  \BibitemShut {NoStop}%
\bibitem [{\citenamefont {Gallego}\ and\ \citenamefont
  {Serone}(2009{\natexlab{a}})}]{Gallego:2008qi}%
  \BibitemOpen
  \bibfield  {author} {\bibinfo {author} {\bibfnamefont {D.}~\bibnamefont
  {Gallego}}\ and\ \bibinfo {author} {\bibfnamefont {M.}~\bibnamefont
  {Serone}},\ }\href {\doibase 10.1088/1126-6708/2009/01/056} {\bibfield
  {journal} {\bibinfo  {journal} {JHEP}\ }\textbf {\bibinfo {volume} {0901}},\
  \bibinfo {pages} {056} (\bibinfo {year} {2009}{\natexlab{a}})},\ \Eprint
  {http://arxiv.org/abs/0812.0369} {arXiv:0812.0369 [hep-th]} \BibitemShut
  {NoStop}%
\bibitem [{\citenamefont {Gallego}\ and\ \citenamefont
  {Serone}(2009{\natexlab{b}})}]{Gallego:2009px}%
  \BibitemOpen
  \bibfield  {author} {\bibinfo {author} {\bibfnamefont {D.}~\bibnamefont
  {Gallego}}\ and\ \bibinfo {author} {\bibfnamefont {M.}~\bibnamefont
  {Serone}},\ }\href {\doibase 10.1088/1126-6708/2009/06/057} {\bibfield
  {journal} {\bibinfo  {journal} {JHEP}\ }\textbf {\bibinfo {volume} {0906}},\
  \bibinfo {pages} {057} (\bibinfo {year} {2009}{\natexlab{b}})},\ \Eprint
  {http://arxiv.org/abs/0904.2537} {arXiv:0904.2537 [hep-th]} \BibitemShut
  {NoStop}%
\bibitem [{\citenamefont {Rummel}\ and\ \citenamefont
  {Westphal}(2012)}]{Rummel:2011cd}%
  \BibitemOpen
  \bibfield  {author} {\bibinfo {author} {\bibfnamefont {M.}~\bibnamefont
  {Rummel}}\ and\ \bibinfo {author} {\bibfnamefont {A.}~\bibnamefont
  {Westphal}},\ }\href {\doibase 10.1007/JHEP01(2012)020} {\bibfield  {journal}
  {\bibinfo  {journal} {JHEP}\ }\textbf {\bibinfo {volume} {1201}},\ \bibinfo
  {pages} {020} (\bibinfo {year} {2012})},\ \Eprint
  {http://arxiv.org/abs/1107.2115} {arXiv:1107.2115 [hep-th]} \BibitemShut
  {NoStop}%
\bibitem [{\citenamefont {Louis}\ \emph {et~al.}(2012)\citenamefont {Louis},
  \citenamefont {Rummel}, \citenamefont {Valandro},\ and\ \citenamefont
  {Westphal}}]{Louis:2012nb}%
  \BibitemOpen
  \bibfield  {author} {\bibinfo {author} {\bibfnamefont {J.}~\bibnamefont
  {Louis}}, \bibinfo {author} {\bibfnamefont {M.}~\bibnamefont {Rummel}},
  \bibinfo {author} {\bibfnamefont {R.}~\bibnamefont {Valandro}}, \ and\
  \bibinfo {author} {\bibfnamefont {A.}~\bibnamefont {Westphal}},\ }\href
  {\doibase 10.1007/JHEP10(2012)163} {\bibfield  {journal} {\bibinfo  {journal}
  {JHEP}\ }\textbf {\bibinfo {volume} {1210}},\ \bibinfo {pages} {163}
  (\bibinfo {year} {2012})},\ \Eprint {http://arxiv.org/abs/1208.3208}
  {arXiv:1208.3208 [hep-th]} \BibitemShut {NoStop}%
\bibitem [{\citenamefont {Demirtas}\ \emph
  {et~al.}(2020{\natexlab{a}})\citenamefont {Demirtas}, \citenamefont {Kim},
  \citenamefont {Mcallister},\ and\ \citenamefont {Moritz}}]{Demirtas:2020ffz}%
  \BibitemOpen
  \bibfield  {author} {\bibinfo {author} {\bibfnamefont {M.}~\bibnamefont
  {Demirtas}}, \bibinfo {author} {\bibfnamefont {M.}~\bibnamefont {Kim}},
  \bibinfo {author} {\bibfnamefont {L.}~\bibnamefont {Mcallister}}, \ and\
  \bibinfo {author} {\bibfnamefont {J.}~\bibnamefont {Moritz}},\ }\href@noop {}
  {\  (\bibinfo {year} {2020}{\natexlab{a}})},\ \Eprint
  {http://arxiv.org/abs/2009.03312} {arXiv:2009.03312 [hep-th]} \BibitemShut
  {NoStop}%
\bibitem [{\citenamefont {Conlon}(2008)}]{Conlon:2008}%
  \BibitemOpen
  \bibfield  {author} {\bibinfo {author} {\bibfnamefont {J.~P.}\ \bibnamefont
  {Conlon}},\ }\href {\doibase 10.1088/1126-6708/2008/03/025} {\bibfield
  {journal} {\bibinfo  {journal} {JHEP}\ }\textbf {\bibinfo {volume} {0803}},\
  \bibinfo {pages} {025} (\bibinfo {year} {2008})},\ \Eprint
  {http://arxiv.org/abs/0710.0873} {arXiv:0710.0873 [hep-th]} \BibitemShut
  {NoStop}%
\bibitem [{\citenamefont {Achucarro}\ \emph {et~al.}(2016)\citenamefont
  {Achucarro}, \citenamefont {Ortiz},\ and\ \citenamefont
  {Sousa}}]{Achucarro:2015kja}%
  \BibitemOpen
  \bibfield  {author} {\bibinfo {author} {\bibfnamefont {A.}~\bibnamefont
  {Achucarro}}, \bibinfo {author} {\bibfnamefont {P.}~\bibnamefont {Ortiz}}, \
  and\ \bibinfo {author} {\bibfnamefont {K.}~\bibnamefont {Sousa}},\ }\href
  {\doibase 10.1103/PhysRevD.94.086012} {\bibfield  {journal} {\bibinfo
  {journal} {Phys. Rev.}\ }\textbf {\bibinfo {volume} {D94}},\ \bibinfo {pages}
  {086012} (\bibinfo {year} {2016})},\ \Eprint
  {http://arxiv.org/abs/1510.01273} {arXiv:1510.01273 [hep-th]} \BibitemShut
  {NoStop}%
\bibitem [{\citenamefont {Sousa}\ and\ \citenamefont
  {Ortiz}(2015)}]{Sousa:2014qza}%
  \BibitemOpen
  \bibfield  {author} {\bibinfo {author} {\bibfnamefont {K.}~\bibnamefont
  {Sousa}}\ and\ \bibinfo {author} {\bibfnamefont {P.}~\bibnamefont {Ortiz}},\
  }\href {\doibase 10.1088/1475-7516/2015/02/017} {\bibfield  {journal}
  {\bibinfo  {journal} {JCAP}\ }\textbf {\bibinfo {volume} {1502}},\ \bibinfo
  {pages} {017} (\bibinfo {year} {2015})},\ \Eprint
  {http://arxiv.org/abs/1408.6521} {arXiv:1408.6521 [hep-th]} \BibitemShut
  {NoStop}%
\bibitem [{\citenamefont {Hebecker}\ and\ \citenamefont
  {March-Russell}(2007)}]{Hebecker:2006bn}%
  \BibitemOpen
  \bibfield  {author} {\bibinfo {author} {\bibfnamefont {A.}~\bibnamefont
  {Hebecker}}\ and\ \bibinfo {author} {\bibfnamefont {J.}~\bibnamefont
  {March-Russell}},\ }\href {\doibase 10.1016/j.nuclphysb.2007.05.003}
  {\bibfield  {journal} {\bibinfo  {journal} {Nucl. Phys. B}\ }\textbf
  {\bibinfo {volume} {781}},\ \bibinfo {pages} {99} (\bibinfo {year} {2007})},\
  \Eprint {http://arxiv.org/abs/hep-th/0607120} {arXiv:hep-th/0607120}
  \BibitemShut {NoStop}%
\bibitem [{\citenamefont {Bena}\ \emph {et~al.}(2019)\citenamefont {Bena},
  \citenamefont {Dudas}, \citenamefont {Gra\~na},\ and\ \citenamefont
  {Lust}}]{Bena:2018fqc}%
  \BibitemOpen
  \bibfield  {author} {\bibinfo {author} {\bibfnamefont {I.}~\bibnamefont
  {Bena}}, \bibinfo {author} {\bibfnamefont {E.}~\bibnamefont {Dudas}},
  \bibinfo {author} {\bibfnamefont {M.}~\bibnamefont {Gra\~na}}, \ and\
  \bibinfo {author} {\bibfnamefont {S.}~\bibnamefont {Lust}},\ }\href {\doibase
  10.1002/prop.201800100} {\bibfield  {journal} {\bibinfo  {journal} {Fortsch.
  Phys.}\ }\textbf {\bibinfo {volume} {67}},\ \bibinfo {pages} {1800100}
  (\bibinfo {year} {2019})},\ \Eprint {http://arxiv.org/abs/1809.06861}
  {arXiv:1809.06861 [hep-th]} \BibitemShut {NoStop}%
\bibitem [{\citenamefont {Demirtas}\ \emph
  {et~al.}(2020{\natexlab{b}})\citenamefont {Demirtas}, \citenamefont {Kim},
  \citenamefont {McAllister},\ and\ \citenamefont {Moritz}}]{Demirtas:2019sip}%
  \BibitemOpen
  \bibfield  {author} {\bibinfo {author} {\bibfnamefont {M.}~\bibnamefont
  {Demirtas}}, \bibinfo {author} {\bibfnamefont {M.}~\bibnamefont {Kim}},
  \bibinfo {author} {\bibfnamefont {L.}~\bibnamefont {McAllister}}, \ and\
  \bibinfo {author} {\bibfnamefont {J.}~\bibnamefont {Moritz}},\ }\href
  {\doibase 10.1103/PhysRevLett.124.211603} {\bibfield  {journal} {\bibinfo
  {journal} {Phys. Rev. Lett.}\ }\textbf {\bibinfo {volume} {124}},\ \bibinfo
  {pages} {211603} (\bibinfo {year} {2020}{\natexlab{b}})},\ \Eprint
  {http://arxiv.org/abs/1912.10047} {arXiv:1912.10047 [hep-th]} \BibitemShut
  {NoStop}%
\bibitem [{\citenamefont {Braun}\ and\ \citenamefont
  {Valandro}(2021)}]{Braun:2020jrx}%
  \BibitemOpen
  \bibfield  {author} {\bibinfo {author} {\bibfnamefont {A.~P.}\ \bibnamefont
  {Braun}}\ and\ \bibinfo {author} {\bibfnamefont {R.}~\bibnamefont
  {Valandro}},\ }\href {\doibase 10.1007/JHEP01(2021)207} {\bibfield  {journal}
  {\bibinfo  {journal} {JHEP}\ }\textbf {\bibinfo {volume} {01}},\ \bibinfo
  {pages} {207} (\bibinfo {year} {2021})},\ \Eprint
  {http://arxiv.org/abs/2009.11873} {arXiv:2009.11873 [hep-th]} \BibitemShut
  {NoStop}%
\bibitem [{\citenamefont {Crin\`o}\ \emph {et~al.}(2020)\citenamefont
  {Crin\`o}, \citenamefont {Quevedo},\ and\ \citenamefont
  {Valandro}}]{Crino:2020qwk}%
  \BibitemOpen
  \bibfield  {author} {\bibinfo {author} {\bibfnamefont {C.}~\bibnamefont
  {Crin\`o}}, \bibinfo {author} {\bibfnamefont {F.}~\bibnamefont {Quevedo}}, \
  and\ \bibinfo {author} {\bibfnamefont {R.}~\bibnamefont {Valandro}},\
  }\href@noop {} {\  (\bibinfo {year} {2020})},\ \Eprint
  {http://arxiv.org/abs/2010.15903} {arXiv:2010.15903 [hep-th]} \BibitemShut
  {NoStop}%
\bibitem [{\citenamefont {Bena}\ \emph {et~al.}(2020)\citenamefont {Bena},
  \citenamefont {Bl\r{a}b\"ack}, \citenamefont {Gra\~na},\ and\ \citenamefont
  {L\"ust}}]{Bena:2020xrh}%
  \BibitemOpen
  \bibfield  {author} {\bibinfo {author} {\bibfnamefont {I.}~\bibnamefont
  {Bena}}, \bibinfo {author} {\bibfnamefont {J.}~\bibnamefont {Bl\r{a}b\"ack}},
  \bibinfo {author} {\bibfnamefont {M.}~\bibnamefont {Gra\~na}}, \ and\
  \bibinfo {author} {\bibfnamefont {S.}~\bibnamefont {L\"ust}},\ }\href@noop {}
  {\  (\bibinfo {year} {2020})},\ \Eprint {http://arxiv.org/abs/2010.10519}
  {arXiv:2010.10519 [hep-th]} \BibitemShut {NoStop}%
\bibitem [{\citenamefont {Bena}\ \emph {et~al.}(2021)\citenamefont {Bena},
  \citenamefont {Bl\r{a}b\"ack}, \citenamefont {Gra\~na},\ and\ \citenamefont
  {L\"ust}}]{Bena:2021wyr}%
  \BibitemOpen
  \bibfield  {author} {\bibinfo {author} {\bibfnamefont {I.}~\bibnamefont
  {Bena}}, \bibinfo {author} {\bibfnamefont {J.}~\bibnamefont {Bl\r{a}b\"ack}},
  \bibinfo {author} {\bibfnamefont {M.}~\bibnamefont {Gra\~na}}, \ and\
  \bibinfo {author} {\bibfnamefont {S.}~\bibnamefont {L\"ust}},\ }\href@noop {}
  {\  (\bibinfo {year} {2021})},\ \Eprint {http://arxiv.org/abs/2103.03250}
  {arXiv:2103.03250 [hep-th]} \BibitemShut {NoStop}%
\bibitem [{\citenamefont {\'Alvarez-Garc\'\i{}a}\ \emph
  {et~al.}(2020)\citenamefont {\'Alvarez-Garc\'\i{}a}, \citenamefont
  {Blumenhagen}, \citenamefont {Brinkmann},\ and\ \citenamefont
  {Schlechter}}]{Blumenhagen:2020ire}%
  \BibitemOpen
  \bibfield  {author} {\bibinfo {author} {\bibfnamefont {R.}~\bibnamefont
  {\'Alvarez-Garc\'\i{}a}}, \bibinfo {author} {\bibfnamefont {R.}~\bibnamefont
  {Blumenhagen}}, \bibinfo {author} {\bibfnamefont {M.}~\bibnamefont
  {Brinkmann}}, \ and\ \bibinfo {author} {\bibfnamefont {L.}~\bibnamefont
  {Schlechter}},\ }\href@noop {} {\  (\bibinfo {year} {2020})},\ \Eprint
  {http://arxiv.org/abs/2009.03325} {arXiv:2009.03325 [hep-th]} \BibitemShut
  {NoStop}%
\bibitem [{\citenamefont {Blanco-Pillado}\ \emph {et~al.}(2021)\citenamefont
  {Blanco-Pillado}, \citenamefont {Sousa}, \citenamefont {Urkiola},\ and\
  \citenamefont {Wachter}}]{Blanco-Pillado:2020wjn}%
  \BibitemOpen
  \bibfield  {author} {\bibinfo {author} {\bibfnamefont {J.~J.}\ \bibnamefont
  {Blanco-Pillado}}, \bibinfo {author} {\bibfnamefont {K.}~\bibnamefont
  {Sousa}}, \bibinfo {author} {\bibfnamefont {M.~A.}\ \bibnamefont {Urkiola}},
  \ and\ \bibinfo {author} {\bibfnamefont {J.~M.}\ \bibnamefont {Wachter}},\
  }\href {\doibase 10.1007/JHEP04(2021)149} {\bibfield  {journal} {\bibinfo
  {journal} {JHEP}\ }\textbf {\bibinfo {volume} {04}},\ \bibinfo {pages} {149}
  (\bibinfo {year} {2021})},\ \Eprint {http://arxiv.org/abs/2007.10381}
  {arXiv:2007.10381 [hep-th]} \BibitemShut {NoStop}%
\bibitem [{\citenamefont {Brodie}\ and\ \citenamefont
  {Marsh}(2016)}]{Brodie:2015kza}%
  \BibitemOpen
  \bibfield  {author} {\bibinfo {author} {\bibfnamefont {C.}~\bibnamefont
  {Brodie}}\ and\ \bibinfo {author} {\bibfnamefont {M.~C.~D.}\ \bibnamefont
  {Marsh}},\ }\href {\doibase 10.1007/JHEP01(2016)037} {\bibfield  {journal}
  {\bibinfo  {journal} {JHEP}\ }\textbf {\bibinfo {volume} {01}},\ \bibinfo
  {pages} {037} (\bibinfo {year} {2016})},\ \Eprint
  {http://arxiv.org/abs/1509.06761} {arXiv:1509.06761 [hep-th]} \BibitemShut
  {NoStop}%
\bibitem [{\citenamefont {Marsh}\ and\ \citenamefont
  {Sousa}(2016)}]{Marsh:2015zoa}%
  \BibitemOpen
  \bibfield  {author} {\bibinfo {author} {\bibfnamefont {M.~C.~D.}\
  \bibnamefont {Marsh}}\ and\ \bibinfo {author} {\bibfnamefont
  {K.}~\bibnamefont {Sousa}},\ }\href {\doibase 10.1007/JHEP03(2016)064}
  {\bibfield  {journal} {\bibinfo  {journal} {JHEP}\ }\textbf {\bibinfo
  {volume} {03}},\ \bibinfo {pages} {064} (\bibinfo {year} {2016})},\ \Eprint
  {http://arxiv.org/abs/1512.08549} {arXiv:1512.08549 [hep-th]} \BibitemShut
  {NoStop}%
\bibitem [{\citenamefont {Junghans}(2019)}]{Junghans:2018gdb}%
  \BibitemOpen
  \bibfield  {author} {\bibinfo {author} {\bibfnamefont {D.}~\bibnamefont
  {Junghans}},\ }\href {\doibase 10.1007/JHEP03(2019)150} {\bibfield  {journal}
  {\bibinfo  {journal} {JHEP}\ }\textbf {\bibinfo {volume} {03}},\ \bibinfo
  {pages} {150} (\bibinfo {year} {2019})},\ \Eprint
  {http://arxiv.org/abs/1811.06990} {arXiv:1811.06990 [hep-th]} \BibitemShut
  {NoStop}%
\bibitem [{\citenamefont {Grimm}\ \emph {et~al.}(2020)\citenamefont {Grimm},
  \citenamefont {Li},\ and\ \citenamefont {Valenzuela}}]{Grimm:2019ixq}%
  \BibitemOpen
  \bibfield  {author} {\bibinfo {author} {\bibfnamefont {T.~W.}\ \bibnamefont
  {Grimm}}, \bibinfo {author} {\bibfnamefont {C.}~\bibnamefont {Li}}, \ and\
  \bibinfo {author} {\bibfnamefont {I.}~\bibnamefont {Valenzuela}},\ }\href
  {\doibase 10.1007/JHEP06(2020)009} {\bibfield  {journal} {\bibinfo  {journal}
  {JHEP}\ }\textbf {\bibinfo {volume} {06}},\ \bibinfo {pages} {009} (\bibinfo
  {year} {2020})},\ \Eprint {http://arxiv.org/abs/1910.09549} {arXiv:1910.09549
  [hep-th]} \BibitemShut {NoStop}%
\bibitem [{\citenamefont {Grimm}\ and\ \citenamefont
  {Louis}(2004)}]{Grimm:2004uq}%
  \BibitemOpen
  \bibfield  {author} {\bibinfo {author} {\bibfnamefont {T.~W.}\ \bibnamefont
  {Grimm}}\ and\ \bibinfo {author} {\bibfnamefont {J.}~\bibnamefont {Louis}},\
  }\href {\doibase 10.1016/j.nuclphysb.2004.08.005} {\bibfield  {journal}
  {\bibinfo  {journal} {Nucl.Phys.}\ }\textbf {\bibinfo {volume} {B699}},\
  \bibinfo {pages} {387} (\bibinfo {year} {2004})},\ \Eprint
  {http://arxiv.org/abs/hep-th/0403067} {arXiv:hep-th/0403067 [hep-th]}
  \BibitemShut {NoStop}%
\bibitem [{\citenamefont {Klemm}(2005)}]{Klemm:2005tw}%
  \BibitemOpen
  \bibfield  {author} {\bibinfo {author} {\bibfnamefont {A.}~\bibnamefont
  {Klemm}},\ }\bibfield  {booktitle} {\emph {\bibinfo {booktitle}
  {{Proceedings, RTN Winter School on Strings, Supergravity and Gauge Theories:
  Trieste, Italy, January 31 - February 4, 2005}}},\ }\href {\doibase
  10.22323/1.019.0002} {\bibfield  {journal} {\bibinfo  {journal} {PoS}\
  }\textbf {\bibinfo {volume} {RTN2005}},\ \bibinfo {pages} {002} (\bibinfo
  {year} {2005})}\BibitemShut {NoStop}%
\bibitem [{\citenamefont {Hosono}\ \emph {et~al.}(1995)\citenamefont {Hosono},
  \citenamefont {Klemm}, \citenamefont {Theisen},\ and\ \citenamefont
  {Yau}}]{Hosono:1994ax}%
  \BibitemOpen
  \bibfield  {author} {\bibinfo {author} {\bibfnamefont {S.}~\bibnamefont
  {Hosono}}, \bibinfo {author} {\bibfnamefont {A.}~\bibnamefont {Klemm}},
  \bibinfo {author} {\bibfnamefont {S.}~\bibnamefont {Theisen}}, \ and\
  \bibinfo {author} {\bibfnamefont {S.-T.}\ \bibnamefont {Yau}},\ }\href
  {\doibase 10.1016/0550-3213(94)00440-P} {\bibfield  {journal} {\bibinfo
  {journal} {Nucl. Phys.}\ }\textbf {\bibinfo {volume} {B433}},\ \bibinfo
  {pages} {501} (\bibinfo {year} {1995})},\ \bibinfo {note} {[,545(1994);
  AMS/IP Stud. Adv. Math.1,545(1996)]},\ \Eprint
  {http://arxiv.org/abs/hep-th/9406055} {arXiv:hep-th/9406055 [hep-th]}
  \BibitemShut {NoStop}%
\bibitem [{\citenamefont {Gukov}\ \emph {et~al.}(2000)\citenamefont {Gukov},
  \citenamefont {Vafa},\ and\ \citenamefont {Witten}}]{Gukov:1999ya}%
  \BibitemOpen
  \bibfield  {author} {\bibinfo {author} {\bibfnamefont {S.}~\bibnamefont
  {Gukov}}, \bibinfo {author} {\bibfnamefont {C.}~\bibnamefont {Vafa}}, \ and\
  \bibinfo {author} {\bibfnamefont {E.}~\bibnamefont {Witten}},\ }\href
  {\doibase 10.1016/S0550-3213(00)00373-4} {\bibfield  {journal} {\bibinfo
  {journal} {Nucl.Phys.}\ }\textbf {\bibinfo {volume} {B584}},\ \bibinfo
  {pages} {69} (\bibinfo {year} {2000})},\ \Eprint
  {http://arxiv.org/abs/hep-th/9906070} {arXiv:hep-th/9906070 [hep-th]}
  \BibitemShut {NoStop}%
\bibitem [{\citenamefont {Candelas}\ \emph
  {et~al.}(1994{\natexlab{a}})\citenamefont {Candelas}, \citenamefont
  {De~La~Ossa}, \citenamefont {Font}, \citenamefont {Katz},\ and\ \citenamefont
  {Morrison}}]{Candelas:1993dm}%
  \BibitemOpen
  \bibfield  {author} {\bibinfo {author} {\bibfnamefont {P.}~\bibnamefont
  {Candelas}}, \bibinfo {author} {\bibfnamefont {X.}~\bibnamefont
  {De~La~Ossa}}, \bibinfo {author} {\bibfnamefont {A.}~\bibnamefont {Font}},
  \bibinfo {author} {\bibfnamefont {S.~H.}\ \bibnamefont {Katz}}, \ and\
  \bibinfo {author} {\bibfnamefont {D.~R.}\ \bibnamefont {Morrison}},\ }\href
  {\doibase 10.1016/0550-3213(94)90322-0} {\bibfield  {journal} {\bibinfo
  {journal} {Nucl.Phys.}\ }\textbf {\bibinfo {volume} {B416}},\ \bibinfo
  {pages} {481} (\bibinfo {year} {1994}{\natexlab{a}})},\ \Eprint
  {http://arxiv.org/abs/hep-th/9308083} {arXiv:hep-th/9308083 [hep-th]}
  \BibitemShut {NoStop}%
\bibitem [{\citenamefont {Mayr}(2001)}]{Mayr:2000as}%
  \BibitemOpen
  \bibfield  {author} {\bibinfo {author} {\bibfnamefont {P.}~\bibnamefont
  {Mayr}},\ }\href {\doibase 10.1088/1126-6708/2001/01/018} {\bibfield
  {journal} {\bibinfo  {journal} {JHEP}\ }\textbf {\bibinfo {volume} {01}},\
  \bibinfo {pages} {018} (\bibinfo {year} {2001})},\ \Eprint
  {http://arxiv.org/abs/hep-th/0010223} {arXiv:hep-th/0010223} \BibitemShut
  {NoStop}%
\bibitem [{\citenamefont {Achucarro}\ and\ \citenamefont
  {Sousa}(2008)}]{Achucarro:2007qa}%
  \BibitemOpen
  \bibfield  {author} {\bibinfo {author} {\bibfnamefont {A.}~\bibnamefont
  {Achucarro}}\ and\ \bibinfo {author} {\bibfnamefont {K.}~\bibnamefont
  {Sousa}},\ }\href {\doibase 10.1088/1126-6708/2008/03/002} {\bibfield
  {journal} {\bibinfo  {journal} {JHEP}\ }\textbf {\bibinfo {volume} {0803}},\
  \bibinfo {pages} {002} (\bibinfo {year} {2008})},\ \Eprint
  {http://arxiv.org/abs/0712.3460} {arXiv:0712.3460 [hep-th]} \BibitemShut
  {NoStop}%
\bibitem [{\citenamefont {Achucarro}\ \emph
  {et~al.}(2008{\natexlab{a}})\citenamefont {Achucarro}, \citenamefont
  {Hardeman},\ and\ \citenamefont {Sousa}}]{Achucarro:2008sy}%
  \BibitemOpen
  \bibfield  {author} {\bibinfo {author} {\bibfnamefont {A.}~\bibnamefont
  {Achucarro}}, \bibinfo {author} {\bibfnamefont {S.}~\bibnamefont {Hardeman}},
  \ and\ \bibinfo {author} {\bibfnamefont {K.}~\bibnamefont {Sousa}},\ }\href
  {\doibase 10.1103/PhysRevD.78.101901} {\bibfield  {journal} {\bibinfo
  {journal} {Phys.Rev.}\ }\textbf {\bibinfo {volume} {D78}},\ \bibinfo {pages}
  {101901(R)} (\bibinfo {year} {2008}{\natexlab{a}})},\ \Eprint
  {http://arxiv.org/abs/0806.4364} {arXiv:0806.4364 [hep-th]} \BibitemShut
  {NoStop}%
\bibitem [{\citenamefont {Achucarro}\ \emph
  {et~al.}(2008{\natexlab{b}})\citenamefont {Achucarro}, \citenamefont
  {Hardeman},\ and\ \citenamefont {Sousa}}]{Achucarro:2008fk}%
  \BibitemOpen
  \bibfield  {author} {\bibinfo {author} {\bibfnamefont {A.}~\bibnamefont
  {Achucarro}}, \bibinfo {author} {\bibfnamefont {S.}~\bibnamefont {Hardeman}},
  \ and\ \bibinfo {author} {\bibfnamefont {K.}~\bibnamefont {Sousa}},\ }\href
  {\doibase 10.1088/1126-6708/2008/11/003} {\bibfield  {journal} {\bibinfo
  {journal} {JHEP}\ }\textbf {\bibinfo {volume} {0811}},\ \bibinfo {pages}
  {003} (\bibinfo {year} {2008}{\natexlab{b}})},\ \Eprint
  {http://arxiv.org/abs/0809.1441} {arXiv:0809.1441 [hep-th]} \BibitemShut
  {NoStop}%
\bibitem [{\citenamefont {Cremmer}\ \emph {et~al.}(1985)\citenamefont
  {Cremmer}, \citenamefont {Kounnas}, \citenamefont {Van~Proeyen},
  \citenamefont {Derendinger}, \citenamefont {Ferrara}, \citenamefont
  {de~Wit},\ and\ \citenamefont {Girardello}}]{Cremmer:1984hj}%
  \BibitemOpen
  \bibfield  {author} {\bibinfo {author} {\bibfnamefont {E.}~\bibnamefont
  {Cremmer}}, \bibinfo {author} {\bibfnamefont {C.}~\bibnamefont {Kounnas}},
  \bibinfo {author} {\bibfnamefont {A.}~\bibnamefont {Van~Proeyen}}, \bibinfo
  {author} {\bibfnamefont {J.}~\bibnamefont {Derendinger}}, \bibinfo {author}
  {\bibfnamefont {S.}~\bibnamefont {Ferrara}}, \bibinfo {author} {\bibfnamefont
  {B.}~\bibnamefont {de~Wit}}, \ and\ \bibinfo {author} {\bibfnamefont
  {L.}~\bibnamefont {Girardello}},\ }\href {\doibase
  10.1016/0550-3213(85)90488-2} {\bibfield  {journal} {\bibinfo  {journal}
  {Nucl. Phys. B}\ }\textbf {\bibinfo {volume} {250}},\ \bibinfo {pages} {385}
  (\bibinfo {year} {1985})}\BibitemShut {NoStop}%
\bibitem [{\citenamefont {Farquet}\ and\ \citenamefont
  {Scrucca}(2012)}]{Farquet:2012cs}%
  \BibitemOpen
  \bibfield  {author} {\bibinfo {author} {\bibfnamefont {D.}~\bibnamefont
  {Farquet}}\ and\ \bibinfo {author} {\bibfnamefont {C.~A.}\ \bibnamefont
  {Scrucca}},\ }\href {\doibase 10.1007/JHEP09(2012)025} {\bibfield  {journal}
  {\bibinfo  {journal} {JHEP}\ }\textbf {\bibinfo {volume} {09}},\ \bibinfo
  {pages} {025} (\bibinfo {year} {2012})},\ \Eprint
  {http://arxiv.org/abs/1205.5728} {arXiv:1205.5728 [hep-th]} \BibitemShut
  {NoStop}%
\bibitem [{\citenamefont {Sousa}(2012)}]{Sousa:2012nvn}%
  \BibitemOpen
  \bibfield  {author} {\bibinfo {author} {\bibfnamefont {K.}~\bibnamefont
  {Sousa}},\ }\emph {\bibinfo {title} {{Consistent Supersymetric Decoupling in
  Cosmology}}},\ \href@noop {} {Ph.D. thesis},\ \bibinfo  {school} {Leiden U.}
  (\bibinfo {year} {2012})\BibitemShut {NoStop}%
\bibitem [{\citenamefont {Denef}\ and\ \citenamefont
  {Douglas}(2004)}]{Denef:2004ze}%
  \BibitemOpen
  \bibfield  {author} {\bibinfo {author} {\bibfnamefont {F.}~\bibnamefont
  {Denef}}\ and\ \bibinfo {author} {\bibfnamefont {M.~R.}\ \bibnamefont
  {Douglas}},\ }\href {\doibase 10.1088/1126-6708/2004/05/072} {\bibfield
  {journal} {\bibinfo  {journal} {JHEP}\ }\textbf {\bibinfo {volume} {0405}},\
  \bibinfo {pages} {072} (\bibinfo {year} {2004})},\ \Eprint
  {http://arxiv.org/abs/hep-th/0404116} {arXiv:hep-th/0404116 [hep-th]}
  \BibitemShut {NoStop}%
\bibitem [{\citenamefont {Candelas}\ \emph
  {et~al.}(1994{\natexlab{b}})\citenamefont {Candelas}, \citenamefont {Font},
  \citenamefont {Katz},\ and\ \citenamefont {Morrison}}]{Candelas:1994hw}%
  \BibitemOpen
  \bibfield  {author} {\bibinfo {author} {\bibfnamefont {P.}~\bibnamefont
  {Candelas}}, \bibinfo {author} {\bibfnamefont {A.}~\bibnamefont {Font}},
  \bibinfo {author} {\bibfnamefont {S.~H.}\ \bibnamefont {Katz}}, \ and\
  \bibinfo {author} {\bibfnamefont {D.~R.}\ \bibnamefont {Morrison}},\ }\href
  {\doibase 10.1016/0550-3213(94)90155-4} {\bibfield  {journal} {\bibinfo
  {journal} {Nucl.Phys.}\ }\textbf {\bibinfo {volume} {B429}},\ \bibinfo
  {pages} {626} (\bibinfo {year} {1994}{\natexlab{b}})},\ \Eprint
  {http://arxiv.org/abs/hep-th/9403187} {arXiv:hep-th/9403187 [hep-th]}
  \BibitemShut {NoStop}%
\bibitem [{\citenamefont {Cicoli}\ \emph {et~al.}(2014)\citenamefont {Cicoli},
  \citenamefont {Klevers}, \citenamefont {Krippendorf}, \citenamefont
  {Mayrhofer}, \citenamefont {Quevedo} \emph {et~al.}}]{Cicoli:2013cha}%
  \BibitemOpen
  \bibfield  {author} {\bibinfo {author} {\bibfnamefont {M.}~\bibnamefont
  {Cicoli}}, \bibinfo {author} {\bibfnamefont {D.}~\bibnamefont {Klevers}},
  \bibinfo {author} {\bibfnamefont {S.}~\bibnamefont {Krippendorf}}, \bibinfo
  {author} {\bibfnamefont {C.}~\bibnamefont {Mayrhofer}}, \bibinfo {author}
  {\bibfnamefont {F.}~\bibnamefont {Quevedo}},  \emph {et~al.},\ }\href
  {\doibase 10.1007/JHEP05(2014)001} {\bibfield  {journal} {\bibinfo  {journal}
  {JHEP}\ }\textbf {\bibinfo {volume} {1405}},\ \bibinfo {pages} {001}
  (\bibinfo {year} {2014})},\ \Eprint {http://arxiv.org/abs/1312.0014}
  {arXiv:1312.0014 [hep-th]} \BibitemShut {NoStop}%
\bibitem [{\citenamefont {Bates}\ \emph {et~al.}(2018)\citenamefont {Bates},
  \citenamefont {Brake},\ and\ \citenamefont {Niemerg}}]{bates2018paramotopy}%
  \BibitemOpen
  \bibfield  {author} {\bibinfo {author} {\bibfnamefont {D.}~\bibnamefont
  {Bates}}, \bibinfo {author} {\bibfnamefont {D.}~\bibnamefont {Brake}}, \ and\
  \bibinfo {author} {\bibfnamefont {M.}~\bibnamefont {Niemerg}},\ }in\
  \href@noop {} {\emph {\bibinfo {booktitle} {International Congress on
  Mathematical Software}}}\ (\bibinfo {organization} {Springer},\ \bibinfo
  {year} {2018})\ pp.\ \bibinfo {pages} {28--35}\BibitemShut {NoStop}%
\bibitem [{\citenamefont {Sommese}\ and\ \citenamefont {Wampler}(1996)}]{SW96}%
  \BibitemOpen
  \bibfield  {author} {\bibinfo {author} {\bibfnamefont {A.~J.}\ \bibnamefont
  {Sommese}}\ and\ \bibinfo {author} {\bibfnamefont {C.~W.}\ \bibnamefont
  {Wampler}},\ }\bibfield  {booktitle} {\emph {\bibinfo {booktitle} {The
  Mathematics of Numerical Analysis}},\ }\href@noop {} {\ \bibinfo {series}
  {Lectures in applied mathematics},\ \textbf {\bibinfo {volume} {32}},\
  \bibinfo {pages} {749} (\bibinfo {year} {1996})}\BibitemShut {NoStop}%
\bibitem [{\citenamefont {Sommese}\ and\ \citenamefont {Wampler}(2005)}]{NSSP}%
  \BibitemOpen
  \bibfield  {author} {\bibinfo {author} {\bibfnamefont {A.~J.}\ \bibnamefont
  {Sommese}}\ and\ \bibinfo {author} {\bibfnamefont {C.~W.}\ \bibnamefont
  {Wampler}},\ }\href@noop {} {\emph {\bibinfo {title} {The Numerical Solution
  of Systems of Polynomials Arising in Engineering and Science}}}\ (\bibinfo
  {publisher} {World Scientific},\ \bibinfo {year} {2005})\BibitemShut
  {NoStop}%
\bibitem [{\citenamefont {Denef}\ \emph {et~al.}(2004)\citenamefont {Denef},
  \citenamefont {Douglas},\ and\ \citenamefont {Florea}}]{Denef:2004dm}%
  \BibitemOpen
  \bibfield  {author} {\bibinfo {author} {\bibfnamefont {F.}~\bibnamefont
  {Denef}}, \bibinfo {author} {\bibfnamefont {M.~R.}\ \bibnamefont {Douglas}},
  \ and\ \bibinfo {author} {\bibfnamefont {B.}~\bibnamefont {Florea}},\ }\href
  {\doibase 10.1088/1126-6708/2004/06/034} {\bibfield  {journal} {\bibinfo
  {journal} {JHEP}\ }\textbf {\bibinfo {volume} {0406}},\ \bibinfo {pages}
  {034} (\bibinfo {year} {2004})},\ \Eprint
  {http://arxiv.org/abs/hep-th/0404257} {arXiv:hep-th/0404257 [hep-th]}
  \BibitemShut {NoStop}%
\bibitem [{\citenamefont {Lust}\ \emph {et~al.}(2006)\citenamefont {Lust},
  \citenamefont {Mayr}, \citenamefont {Reffert},\ and\ \citenamefont
  {Stieberger}}]{Lust:2005bd}%
  \BibitemOpen
  \bibfield  {author} {\bibinfo {author} {\bibfnamefont {D.}~\bibnamefont
  {Lust}}, \bibinfo {author} {\bibfnamefont {P.}~\bibnamefont {Mayr}}, \bibinfo
  {author} {\bibfnamefont {S.}~\bibnamefont {Reffert}}, \ and\ \bibinfo
  {author} {\bibfnamefont {S.}~\bibnamefont {Stieberger}},\ }\href {\doibase
  10.1016/j.nuclphysb.2005.09.011} {\bibfield  {journal} {\bibinfo  {journal}
  {Nucl. Phys. B}\ }\textbf {\bibinfo {volume} {732}},\ \bibinfo {pages} {243}
  (\bibinfo {year} {2006})},\ \Eprint {http://arxiv.org/abs/hep-th/0501139}
  {arXiv:hep-th/0501139} \BibitemShut {NoStop}%
\bibitem [{\citenamefont {Collinucci}\ \emph {et~al.}(2009)\citenamefont
  {Collinucci}, \citenamefont {Denef},\ and\ \citenamefont
  {Esole}}]{Collinucci:2008pf}%
  \BibitemOpen
  \bibfield  {author} {\bibinfo {author} {\bibfnamefont {A.}~\bibnamefont
  {Collinucci}}, \bibinfo {author} {\bibfnamefont {F.}~\bibnamefont {Denef}}, \
  and\ \bibinfo {author} {\bibfnamefont {M.}~\bibnamefont {Esole}},\ }\href
  {\doibase 10.1088/1126-6708/2009/02/005} {\bibfield  {journal} {\bibinfo
  {journal} {JHEP}\ }\textbf {\bibinfo {volume} {02}},\ \bibinfo {pages} {005}
  (\bibinfo {year} {2009})},\ \Eprint {http://arxiv.org/abs/0805.1573}
  {arXiv:0805.1573 [hep-th]} \BibitemShut {NoStop}%
\bibitem [{\citenamefont {Alim}\ \emph {et~al.}(2010)\citenamefont {Alim},
  \citenamefont {Hecht}, \citenamefont {Jockers}, \citenamefont {Mayr},
  \citenamefont {Mertens},\ and\ \citenamefont {Soroush}}]{Alim:2009bx}%
  \BibitemOpen
  \bibfield  {author} {\bibinfo {author} {\bibfnamefont {M.}~\bibnamefont
  {Alim}}, \bibinfo {author} {\bibfnamefont {M.}~\bibnamefont {Hecht}},
  \bibinfo {author} {\bibfnamefont {H.}~\bibnamefont {Jockers}}, \bibinfo
  {author} {\bibfnamefont {P.}~\bibnamefont {Mayr}}, \bibinfo {author}
  {\bibfnamefont {A.}~\bibnamefont {Mertens}}, \ and\ \bibinfo {author}
  {\bibfnamefont {M.}~\bibnamefont {Soroush}},\ }\href {\doibase
  10.1016/j.nuclphysb.2010.06.017} {\bibfield  {journal} {\bibinfo  {journal}
  {Nucl. Phys. B}\ }\textbf {\bibinfo {volume} {841}},\ \bibinfo {pages} {303}
  (\bibinfo {year} {2010})},\ \Eprint {http://arxiv.org/abs/0909.1842}
  {arXiv:0909.1842 [hep-th]} \BibitemShut {NoStop}%
\bibitem [{\citenamefont {Honma}\ and\ \citenamefont
  {Otsuka}(2017)}]{Honma:2017uzn}%
  \BibitemOpen
  \bibfield  {author} {\bibinfo {author} {\bibfnamefont {Y.}~\bibnamefont
  {Honma}}\ and\ \bibinfo {author} {\bibfnamefont {H.}~\bibnamefont {Otsuka}},\
  }\href {\doibase 10.1016/j.physletb.2017.09.062} {\bibfield  {journal}
  {\bibinfo  {journal} {Phys. Lett. B}\ }\textbf {\bibinfo {volume} {774}},\
  \bibinfo {pages} {225} (\bibinfo {year} {2017})},\ \Eprint
  {http://arxiv.org/abs/1706.09417} {arXiv:1706.09417 [hep-th]} \BibitemShut
  {NoStop}%
\bibitem [{\citenamefont {Honma}\ and\ \citenamefont
  {Otsuka}(2020)}]{Honma:2019gzp}%
  \BibitemOpen
  \bibfield  {author} {\bibinfo {author} {\bibfnamefont {Y.}~\bibnamefont
  {Honma}}\ and\ \bibinfo {author} {\bibfnamefont {H.}~\bibnamefont {Otsuka}},\
  }\href {\doibase 10.1007/JHEP04(2020)001} {\bibfield  {journal} {\bibinfo
  {journal} {JHEP}\ }\textbf {\bibinfo {volume} {04}},\ \bibinfo {pages} {001}
  (\bibinfo {year} {2020})},\ \Eprint {http://arxiv.org/abs/1910.10725}
  {arXiv:1910.10725 [hep-th]} \BibitemShut {NoStop}%
\bibitem [{\citenamefont {Martinez-Pedrera}\ \emph {et~al.}(2013)\citenamefont
  {Martinez-Pedrera}, \citenamefont {Mehta}, \citenamefont {Rummel},\ and\
  \citenamefont {Westphal}}]{MartinezPedrera:2012rs}%
  \BibitemOpen
  \bibfield  {author} {\bibinfo {author} {\bibfnamefont {D.}~\bibnamefont
  {Martinez-Pedrera}}, \bibinfo {author} {\bibfnamefont {D.}~\bibnamefont
  {Mehta}}, \bibinfo {author} {\bibfnamefont {M.}~\bibnamefont {Rummel}}, \
  and\ \bibinfo {author} {\bibfnamefont {A.}~\bibnamefont {Westphal}},\ }\href
  {\doibase 10.1007/JHEP06(2013)110} {\bibfield  {journal} {\bibinfo  {journal}
  {JHEP}\ }\textbf {\bibinfo {volume} {06}},\ \bibinfo {pages} {110} (\bibinfo
  {year} {2013})},\ \Eprint {http://arxiv.org/abs/1212.4530} {arXiv:1212.4530
  [hep-th]} \BibitemShut {NoStop}%
\bibitem [{\citenamefont {DeWolfe}\ \emph {et~al.}(2005)\citenamefont
  {DeWolfe}, \citenamefont {Giryavets}, \citenamefont {Kachru},\ and\
  \citenamefont {Taylor}}]{DeWolfe:2004ns}%
  \BibitemOpen
  \bibfield  {author} {\bibinfo {author} {\bibfnamefont {O.}~\bibnamefont
  {DeWolfe}}, \bibinfo {author} {\bibfnamefont {A.}~\bibnamefont {Giryavets}},
  \bibinfo {author} {\bibfnamefont {S.}~\bibnamefont {Kachru}}, \ and\ \bibinfo
  {author} {\bibfnamefont {W.}~\bibnamefont {Taylor}},\ }\href {\doibase
  10.1088/1126-6708/2005/02/037} {\bibfield  {journal} {\bibinfo  {journal}
  {JHEP}\ }\textbf {\bibinfo {volume} {0502}},\ \bibinfo {pages} {037}
  (\bibinfo {year} {2005})},\ \Eprint {http://arxiv.org/abs/hep-th/0411061}
  {arXiv:hep-th/0411061 [hep-th]} \BibitemShut {NoStop}%
\bibitem [{\citenamefont {Denef}\ and\ \citenamefont
  {Douglas}(2005)}]{Denef:2004cf}%
  \BibitemOpen
  \bibfield  {author} {\bibinfo {author} {\bibfnamefont {F.}~\bibnamefont
  {Denef}}\ and\ \bibinfo {author} {\bibfnamefont {M.~R.}\ \bibnamefont
  {Douglas}},\ }\href {\doibase 10.1088/1126-6708/2005/03/061} {\bibfield
  {journal} {\bibinfo  {journal} {JHEP}\ }\textbf {\bibinfo {volume} {0503}},\
  \bibinfo {pages} {061} (\bibinfo {year} {2005})},\ \Eprint
  {http://arxiv.org/abs/hep-th/0411183} {arXiv:hep-th/0411183 [hep-th]}
  \BibitemShut {NoStop}%
\bibitem [{\citenamefont {Marsh}\ \emph {et~al.}(2012)\citenamefont {Marsh},
  \citenamefont {McAllister},\ and\ \citenamefont {Wrase}}]{Marsh:2011aa}%
  \BibitemOpen
  \bibfield  {author} {\bibinfo {author} {\bibfnamefont {D.}~\bibnamefont
  {Marsh}}, \bibinfo {author} {\bibfnamefont {L.}~\bibnamefont {McAllister}}, \
  and\ \bibinfo {author} {\bibfnamefont {T.}~\bibnamefont {Wrase}},\ }\href
  {\doibase 10.1007/JHEP03(2012)102} {\bibfield  {journal} {\bibinfo  {journal}
  {JHEP}\ }\textbf {\bibinfo {volume} {1203}},\ \bibinfo {pages} {102}
  (\bibinfo {year} {2012})},\ \Eprint {http://arxiv.org/abs/1112.3034}
  {arXiv:1112.3034 [hep-th]} \BibitemShut {NoStop}%
\end{thebibliography}%

\end{document}